\documentclass[fleqn, usenatbib]{mnras}

\usepackage{newtxtext,newtxmath}
\usepackage[T1]{fontenc}
\usepackage{ae,aecompl}

\usepackage{graphicx}	% Including figure files
\usepackage{amsmath}	% Advanced maths commands
\usepackage{amssymb}	% Extra maths symbols
\usepackage{indentfirst}
\usepackage{hyperref}
\usepackage{booktabs}
\usepackage{cite}
\usepackage[hang,flushmargin]{footmisc}
\usepackage{siunitx}

\newcommand*\diff{\;\mathop{}\!\mathrm{d}}

\title[Understanding the HERA Phase I receiver system]{Understanding the HERA Phase I receiver system with simulations and its impact on the detectability of the EoR delay power spectrum}

\author[Fagnoni et al.]{\normalsize Nicolas Fagnoni$^{1}$\thanks{E-mail: nf323@mrao.cam.ac.uk},
Eloy de Lera Acedo$^{1}$\thanks{E-mail: eloy@mrao.cam.ac.uk},
David R. DeBoer$^{2}$\thanks{E-mail: ddeboer@berkeley.edu},
Zara Abdurashidova$^{2}$,
James E. Aguirre$^{3}$,
Paul Alexander$^{1}$,
\newauthor
\normalsize
Zaki S. Ali$^{2}$,
Yanga Balfour$^{4}$,
Adam P. Beardsley$^{5}$,
Gianni Bernardi$^{4, 6, 7}$,
Tashalee S. Billings$^{3}$,
Judd D. Bowman$^{5}$,
\newauthor
\normalsize
Richard F. Bradley$^{8}$,
Phil Bull$^{9}$,
Jacob Burba$^{10}$,
Chris L. Carilli$^{11}$,
Carina Cheng$^{2}$,
Matt Dexter$^{2}$,
Joshua S. Dillon$^{2}$,
\newauthor
\normalsize
Aaron Ewall-Wice$^{12}$,
Randall Fritz$^{4}$,
Steve R. Furlanetto$^{13}$,
Kingsley Gale-Sides$^{1}$,
Brian Glendenning$^{11}$,
Deepthi Gorthi$^{2}$,
\newauthor
\normalsize
Bradley Greig$^{14}$,
Jasper Grobbelaar$^{4}$,
Ziyaad Halday$^{4}$,
Bryna J. Hazelton$^{15, 16}$,
Jacqueline N. Hewitt$^{12}$,
Jack Hickish$^{2}$,
\newauthor
\normalsize
Daniel C. Jacobs$^{5}$,
Alec Josaitis$^{1}$,
Austin Julius$^{4}$,
Nicholas S. Kern$^{2}$,
Joshua Kerrigan$^{10}$,
Honggeun Kim$^{12}$,
Piyanat Kittiwisit$^{5}$,
\newauthor
\normalsize
Saul A. Kohn$^{3}$,
Matthew Kolopanis$^{5}$,
Adam Lanman$^{10}$,
Paul La Plante$^{3}$,
Telalo Lekalake$^{4}$,
Adrian Liu$^{17}$,
David MacMahon$^{2}$,
\newauthor
\normalsize
Lourence Malan$^{4}$,
Cresshim Malgas$^{4}$,
Matthys Maree$^{4}$,
Zachary E. Martinot$^{3}$,
Eunice Matsetela$^{4}$,
Juan Mena Parra$^{12}$,
\newauthor
\normalsize
Andrei Mesinger$^{18}$,
Mathakane Molewa$^{4}$,
Miguel F. Morales$^{15}$,
Tshegofalang Mosiane$^{4}$,
Abraham R. Neben$^{12}$,
Bojan Nikolic$^{1}$,
\newauthor
\normalsize
Aaron R. Parsons$^{2}$,
Nipanjana Patra$^{2}$,
Samantha Pieterse$^{4}$,
Jonathan C. Pober$^{10}$,
Nima Razavi-Ghods$^{1}$,
James Robnett$^{11}$,
\newauthor
\normalsize
Kathryn Rosie$^{4}$,
Peter Sims$^{10}$,
Craig Smith$^{4}$,
Angelo Syce$^{4}$,
Nithyanandan Thyagarajan$^{5, 11}$,
Peter K. G. Williams$^{19}$,
Haoxuan Zheng$^{12}$.}

\date{Accepted XXX. Received YYY; in original form ZZZ}

\pubyear{2019}

\begin{document}
\label{firstpage}
\pagerange{\pageref{firstpage}--\pageref{lastpage}}
\maketitle

%%%%%%%%%%%%%%%%% ABSTRACT %%%%%%%%%%%%%%%%%%

\begin{abstract}
The detection of the Epoch of Reionization (EoR) delay power spectrum using a "foreground avoidance method" highly depends on the instrument chromaticity. The systematic effects induced by the radio-telescope spread the foreground signal in the delay domain, which contaminates the EoR window theoretically observable. Applied to the Hydrogen Epoch of Reionization Array (HERA), this paper combines detailed electromagnetic and electrical simulations in order to model the chromatic effects of the instrument, and quantify its frequency and time responses. In particular, the effects of the analogue receiver, transmission cables, and mutual coupling are included. These simulations are able to accurately predict the intensity of the reflections occurring in the 150-m cable which links the antenna to the back-end. They also show that electromagnetic waves can propagate from one dish to another one through large sections of the array due to mutual coupling. The simulated system time response is attenuated by a factor $10^{4}$ after a characteristic delay which depends on the size of the array and on the antenna position. Ultimately, the system response is attenuated by a factor $10^{5}$ after 1400 ns because of the reflections in the cable, which corresponds to characterizable ${k_\parallel}$-modes above 0.7 $h\;\rm{Mpc}^{-1}$ at 150 MHz. Thus, this new study shows that the detection of the EoR signal with HERA Phase I will be more challenging than expected. On the other hand, it improves our understanding of the telescope, which is essential to mitigate the instrument chromaticity.
\end{abstract}


%%%%%%%%%%%%%%%%% LIST OF INSTITUTIONS %%%%%%%%%%%%%%%%%%

\begin{flushleft}
\textit{$^{1}$Cavendish Astrophysics, University of Cambridge, Cambridge, UK\\
$^{2}$Department of Astronomy, University of California, Berkeley, CA, US\\
$^{3}$Department of Physics and Astronomy, University of Pennsylvania, Philadelphia, PA, US\\
$^{4}$South African Radio Astronomy Observatory, Cape Town, South Africa\\
$^{5}$School of Earth and Space Exploration, Arizona State University, Tempe, AZ, US\\
$^{6}$Department of Physics and Electronics, Rhodes University, Grahamstown, South Africa\\
$^{7}$INAF-Istituto di Radioastronomia, Bologna, Italy\\
$^{8}$National Radio Astronomy Observatory, Charlottesville, VA, US\\
$^{9}$Queen Mary University of London, London, UK\\
$^{10}$Department of Physics, Brown University, Providence, RI, US\\
$^{11}$National Radio Astronomy Observatory, Socorro, NM, US\\
$^{12}$Department of Physics, Massachusetts Institute of Technology, Cambridge, MA, US\\
$^{13}$Department of Physics and Astronomy, University of California, Los Angeles, CA, US\\
$^{14}$School of Physics, University of Melbourne, Parkville, Australia\\
$^{15}$Department of Physics, University of Washington, Seattle, WA, US\\
$^{16}$eScience Institute, University of Washington, Seattle, WA, US\\
$^{17}$Department of Physics and McGill Space Institute, McGill University, Montreal, Canada\\
$^{18}$Scuola Normale Superiore, Pisa, Italy\\
$^{19}$Harvard-Smithsonian Center for Astrophysics, Cambridge, MA, US}
\end{flushleft}

%%%%%%%%%%%%%%%%% INTRODUCTION %%%%%%%%%%%%%%%%%%

\section{Introduction} \label{sec:1.Intro}

\medskip The Hydrogen Epoch of Reionization Array (HERA) \footnote{https://reionization.org/} is a radio-telescope under development, and dedicated to the study of the early universe. During the Dark Ages, until about 300 million years after the Big Bang, the background emission was dominated by the Cosmic Microwave Background and the 21-cm hydrogen signal. Following the formation of the first stars and galaxies, the neutral hydrogen present in the intergalactic medium started to be heated and ionized by the X-rays and UV radiation emitted. Due to the decrease in the quantity of neutral hydrogen, the 21-cm signal progressively disappeared. The study of the spectral and spatial fluctuations of this signal during the Cosmic Dawn and the EoR is the key to better understand the formation and evolution of the first structures of our universe \citep{McQuinn2007, Morales2010, Pritchard2012, Furlanetto2015}.

\medskip HERA Phase I has been designed to characterize the delay power spectrum of the redshifted hydrogen signal between 100 and 200 MHz \citep{Pober2014, DeBoer2016}. This is equivalent to probing a period between 325 and 915 million years after the Big Bang. This radio-interferometer is being built in the Karoo desert in South Africa. Its final configuration comprises 350 antennas, 320 forming a dense hexagonal core with short baselines, plus 30 outriggers to improve the resolution \citep{Dillon2016}. The high number of redundant baselines is an essential aspect of the telescope and is used to calibrate the system \citep{Liu2010, DeBoer2016, Dillon2018}. Successor of the Precision Array for Probing the Epoch of Reionization (PAPER) \citep{Parsons2010, Ali2015, Pober2015}, HERA Phase I consists of a 14-m diameter parabola and re-uses the dipole feed, the receiver, and the correlator developed for this instrument. This paper focuses on the characterization of the performance of the phase I system, which has been producing data since 2015. As for the phase II, these elements have been redesigned \citep{Fagnoni2020} (Razavi-Ghods et al. 2020 -- in preparation) and are being tested in the desert. The goal is to improve the performance and extend the bandwidth from 50 to 250 MHz, in order to study the Cosmic Dawn and the Reionization, between 115 million and 1.3 billion years after the Big Bang.

\medskip The received signal is the combination of the EoR signal with the foreground. The order of magnitude of the global 21-cm brightness temperature during the EoR is expected to be around 10 mK \citep{Bowman2009}, whereas the temperature of a "cold" patch of sky smoothly decreases from about 1000 K to 100 K, between 100 and 200 MHz. This makes the detection very challenging. Therefore, it is essential to get rid of the contribution from the foreground, either by modelling and subtracting it, or by analysing specific portions of the EoR delay power spectrum which are not affected by the foreground \citep{Chapman2016}. The first method, difficult to achieve in practice, requires a very accurate knowledge of the sky along with the instrument response in order to calibrate the system \citep{Ewall-Wice2017}. For example, this method has been applied by the LOw Frequency ARray (LOFAR) \citep{Harker2010, Chapman2012, VanHaarlem2013}, and by the Murchison Widefield Array (MWA) \citep{Tingay2013, Jacobs2016}. The second option which has been initially preferred for HERA is more straightforward, but also more conservative since the portions of the EoR delay power spectrum significantly contaminated by the foreground are in principle not used to characterize the signal. This foreground avoidance method developed in \citet{Parsons2012a, Parsons2012b} and in \citet{Liu2014} is based on the fact that the foreground is relatively smooth in the plane of the sky and in the frequency domain. On the other hand, the EoR signal varies relatively quickly in the \textit{uv}-plane and along the line of sight. By translating this property into the delay domain, one obtains a very compact foreground delay power spectrum, while the EoR delay power spectrum spreads. This difference can be used to separate the two signals. The effectiveness of this method mainly depends on the smoothness of the instrument response and of the observed sky \citep{Thyagarajan2016}. 

\medskip This paper combines electromagnetic simulations and microwave engineering techniques to better understand and quantify the effects of the telescope on the received signal. This aspect is critical for the success of this experiment. In \citet{Ewall-Wice2016} and \citet{Patra2018}, the instrument response and its impact on the EoR detection were analysed for a single antenna terminated by a fixed impedance. The purpose of this paper is to complete this work, by applying a rigorous method to simulate all the main sources of chromaticity affecting the system in a large array. Two new elements are added: a model of the analogue receiver including the cables, and the effects of mutual coupling between antennas. By studying the system response in reception, this approach allows us to better quantify the chromatic effects in the array and draw new conclusions concerning its ability to detect the EoR delay power spectrum with a technique purely based on the foreground avoidance.  

\medskip Section \ref{sec:2.ForegroundAv} summarizes how the delay power spectrum can be obtained from the output voltages generated by a baseline. We also explain how the foreground contamination can be avoided and how the chromatic effects can affect the detection of the EoR signal. Section \ref{sec:3.Co-simuPara} presents the models and parameters used to simulate the properties of the antenna along with the receiver. The methodology applied to perform these co-simulations and obtain the end-to-end system voltage response is detailed in Section \ref{sec:4.Methodology}. Section \ref{sec:5.RecCplPerf} presents the results and discusses the impact of the receiver and mutual coupling on the performance of the system. These simulations are used to estimate the system noise temperature, as well as to study the propagation and reflection of the voltage signal through the system. To validate these simulations, the results are compared with reflectometry and noise measurements. Lastly, we discuss the implications of these new results on the ability to characterize the EoR delay power spectrum with a direct foreground avoidance method.

%%%%%%%%%%%%%%%%% SECTION 2 %%%%%%%%%%%%%%%%%%

\section{Detection of the EoR delay power spectrum in the "foreground avoidance method"} \label{sec:2.ForegroundAv}

%%%%%%%%%%%%%%%%% SUB-SECTION 2.1 %%%%%%%%%%%%%%%%%%

\subsection{Antenna output voltage and delay power spectrum} \label{sec:2.1.DelSpec}

\medskip HERA is a radio-interferometer which is designed to perform a per-baseline analysis of the EoR signal: each pair of antennas is independent and samples the delay power spectrum associated with a specific spatial scale and for a given frequency band. In this section, we summarize how this spectrum can be derived from the output voltages measured by a baseline. An electric field received by an antenna \textit{i}, and coming from the direction defined by the vector $\mathbfit{s}$ perpendicular to the incident wavefront, is transformed into an analogue voltage signal. Assuming that the system is "linear and time invariant", the total output voltage in the time domain is obtained by convolving the incoming electric field $\bmath{{e_{\rm i}}} \left( {\mathbfit{s},\;t} \right)$ with the impulse voltage time response of the system $\bmath{{h_{\rm i}}} \left( {\mathbfit{s},\;t} \right)$, and by summing the contributions received from all directions. This is equivalent to a multiplication in the frequency domain, where these quantities are respectively denoted $\bmath{{E_{\rm i}}} \left( {\mathbfit{s},\;f} \right)$ and $\bmath{{H_{\rm i}}} \left( {\mathbfit{s},\;f} \right)$. The voltage signal is measured, digitized, and Fourier transformed into the frequency domain. The data are then sent to the correlator \citep{Parsons2008} which computes the visibility associated with each baseline. It can be expressed as a function of the measured voltages, sky brightness, and system responses. For instance, the visibility defined by the antennas 1 and 2 is equal to \citep[Ch. 2-3]{Thomson2017}:
\begin{align}
    {V_{12}}\left( f \right) = \mathop \int\!\!\!\int \nolimits_{4\pi }^{} I\left( {\mathbfit{s}, \;f} \right)A\left( {\mathbfit{s}, \;f} \right)W\left( f \right){\mathrm{e}^{-2\mathrm{j}\pi f{\tau_{\rm 12}}}}\;\diff \Omega \;,
	\label{eq1:SimpVisibility}
\end{align}
with $I\left( {\mathbfit{s}, \;f} \right) = \langle\;{{| {\bmath{{E_{\rm 1}}} \left( {\mathbfit{s}, \;f} \right)}| }^2\rangle}_{\rm T}$ the time-averaged sky intensity, $A\left( {\mathbfit{s}, \;f} \right) = \overline {\bmath{{H_{\rm 1}}} \left( {\mathbfit{s}, \;f} \right)} \boldsymbol{\cdot}\bmath{{H_{\rm 2}}} \left( {\mathbfit{s}, \;f} \right)$ the frequency response of the baseline, $W\left( f \right)$ a window function used to account for the effects of the limited bandwidth of the correlator, ${\tau_{\rm 12}}\left( {\bmath{b}, \;\mathbfit{s}} \right) = {\mathbfit{b} \boldsymbol{\cdot}\mathbfit{s}}/{c}$ the delay of the received signal between the two antennas caused by the baseline length defined by the vector $\mathbfit{b}$, and$\diff \Omega$ the solid angle element. To obtain this equation, it is assumed that signals coming from different radio sources and directions are not coherent, and therefore that their cross-correlation is null. We also disregard the additional random noise voltages, generated for instance by the receiver, in order to focus the analysis on the received radio signal.

\medskip The delay spectrum ${\tilde v_{\rm 12}}\left( \tau  \right)$ is then obtained by taking the inverse Fourier transform of the visibility. The goal is to separate the EoR signal from the foreground in the delay domain, thanks to their spectral differences.
\begin{align}
    {\tilde v_{\rm 12}}\left( \tau  \right) = \mathop \int \int\!\!\!\int \nolimits_{4\pi }^{} I\left( {\mathbfit{s}, \;f} \right)A\left( {\mathbfit{s}, \;f} \right)W\left( f \right){\mathrm{e}^{2\mathrm{j} \pi f\left( {\tau  - {\tau_{\rm 12}}} \right)}}\diff \Omega\;\diff f.
	\label{eq2:delSpectrum}
\end{align}
From the delay spectrum, \citet{Parsons2012a, Parsons2012b} details how to obtain the cosmological delay power spectrum associated with the redshifted hydrogen signal whose frequency is directly related to its epoch of emission. Averaged in a cylindrical volume, the delay power spectrum contains information about the spatial and time evolution of the distribution of the neutral hydrogen. Expressed in K$^{2}$ $(h^{-1}\rm{Mpc})^{3}$, it is is equal to \citep{Thyagarajan2015a}:
\begin{align}
    {P{\rm _d}}\left( {\bmath{{k_ \bot }} ,\;{k_\parallel }} \right) = {\left| {{{\tilde v}_{12}}\left( \tau  \right)} \right|^2}{\left( {\frac{{{\lambda ^2}}}{{2{k{\rm _B}}}}} \right)^2}\left( {\frac{{D^2}{\Delta D}}{{\Delta B}}} \right)\left( {\frac{1}{{\Omega \Delta B}}} \right), 
	\label{eq3:delPoWSpectrum}
\end{align}
with $\lambda $ the wavelength corresponding to the centre of the observation frequency band, $\Delta B$ its bandwidth, ${k{\rm _B}}$ the Boltzmann constant, $D$ the comoving distance perpendicular to the line-of-sight, $\Delta D$ the comoving depth along the line-of-sight associated with $\Delta B$, and $\Omega\Delta B$ a normalization factor equal to:
\begin{align}
	\Omega\Delta B = \mathop \int \nolimits_{\Delta B}^{} \mathop \int\!\!\!\int \nolimits_{4\pi }^{} {\left| {A\left( {\mathbfit{s}, \;f} \right)W\left( f \right)} \right|^2}\diff \Omega \diff f.
	\label{eq4:beamNormFact}
\end{align}

\medskip The delay power spectrum is usually expressed as a function of the wave vector $\mathbfit{k}$ split into two components $\bmath{{k_ \bot }} $ and ${k_\parallel }$: 
\begin{align}
    \bmath{{k_ \bot }}  = \frac{{2\pi \mathbfit{b}}}{{D\lambda }}, \hspace{1cm} k_\parallel = \frac{2\pi f_{\rm 21} H_{\rm p}\left( z \right) \tau}{c\left( 1+z \right)^{2}} = 2\pi\tau \frac{\Delta B}{\Delta D},
	\label{eq5:kParaPerp}
\end{align}
with $f_{21}$ the emission frequency of the spin-flip transition of the neutral hydrogen equal to 1420.4 MHz, and $H_{\rm p}\left( z \right)$ the Hubble parameter expressed as a function of the redshift $z$. $\;\bmath{{k_ \bot }} $ corresponds to the radial or "perpendicular" component of the wave vector. It is related to the diameter of the observation cylinder, in other words the spatial scale of the observed region in the plane of sky, and is limited by the field of view of the instrument and the baseline length. As for ${k_\parallel}$, it is associated with the "depth" of this volume along the line-of-sight, and is a function of the signal delay and bandwidth. The bandwidth can be adjusted thanks to the weighting function $W\left( f \right)$, and in practice is selected in such a way that the cosmological time evolution of the EoR is negligible. Typically, this corresponds to a variation of redshift $\Delta z$ below 0.5. Thus, at 150 MHz or $z=8.5$, a bandwidth of about 8 MHz is suitable \citep{Mesinger2014}. In the next section, we outline the avoidance method used to isolate the EoR delay power spectrum from the foreground.

%%%%%%%%%%%%%%%%% SUB-SECTION 2.2 %%%%%%%%%%%%%%%%%%

\subsection{The importance of the instrument chromaticity in the foreground avoidance method} \label{sec:2.2.InstruChroma}

\medskip Equation \eqref{eq2:delSpectrum} shows that the delay spectrum depends on the observed sky, the choice of the frequency window, and the instrument response. By observing cold patches of sky, it is possible to receive a foreground which is relatively spectrally smooth. This translates into a compact delay power spectrum \citep{Thyagarajan2015a, Thyagarajan2015b}. On the other hand, due to the spectral and spatial fluctuations of the distribution of neutral hydrogen during the Reionization, its delay power spectrum spreads. Thus, at high delays the EoR signal should dominate the foreground in theory. However, the spectral properties of the selected window and the instrument response can also be responsible for the leakage of the foreground at high delays.

\medskip Leakage in the time domain associated with windowing in the frequency domain is a classic problem in signal analysis. A rectangular frequency window would be a bad choice in this context due to its lack of dynamic range ($10^2$ -- $10^3$). \citet{Parsons2012b} and \citet{Thyagarajan2013, Thyagarajan2016} studied this question and concluded that a Blackman-Harris window \citep{Harris1978} can provide satisfactory performance, with a dynamic range of more than $10^6$ for voltage signals in the time domain. Thus, the foreground leakage caused by the window can be kept below the EoR signal, but to the detriment of the sensitivity which is reduced by about 50\%.

\begin{figure}
  \includegraphics[width=0.95\linewidth]{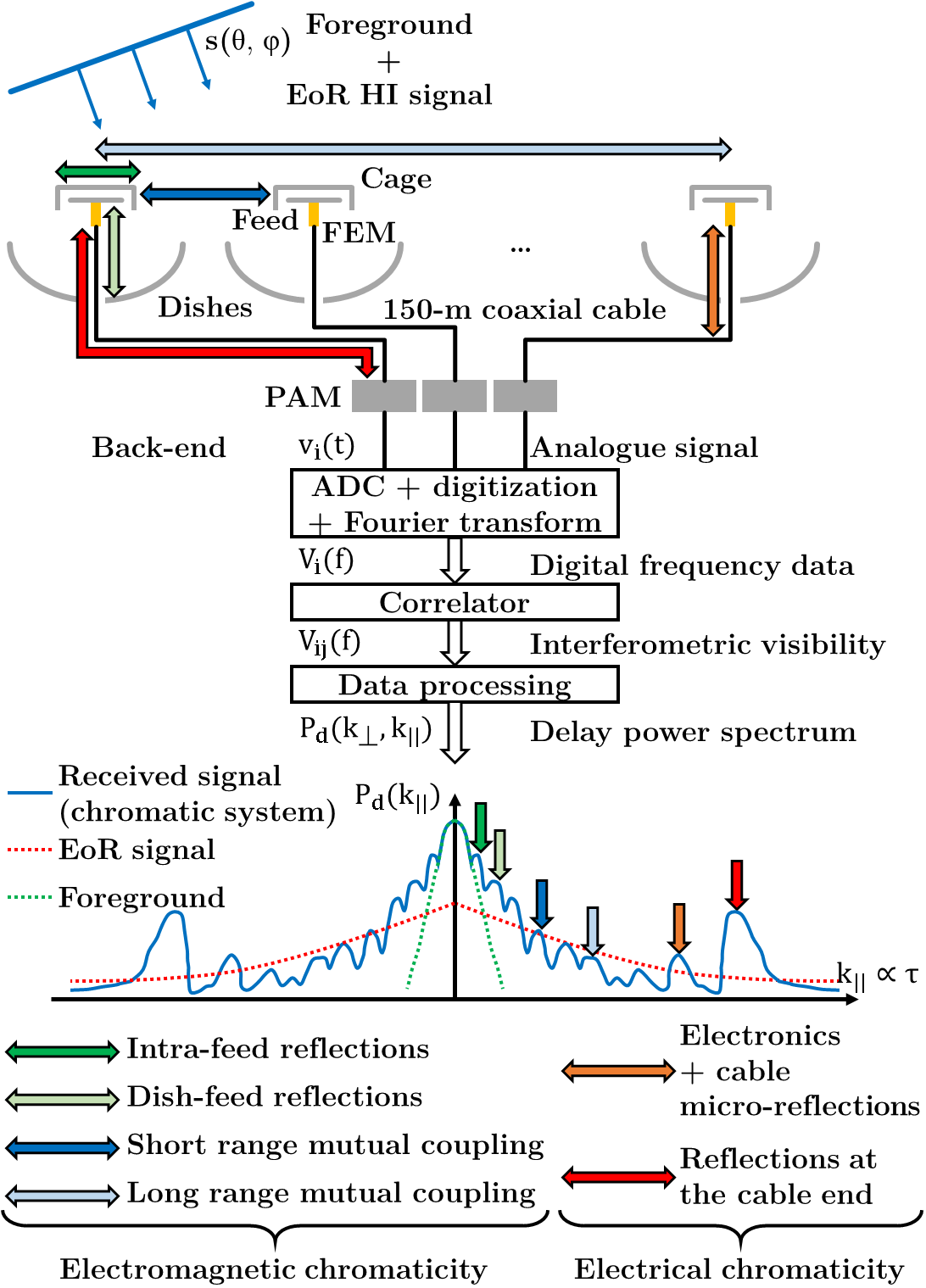}
  \caption{Figure schematically illustrating the HERA receiver system and the main sources of chromaticity affecting the detection of the EoR signal.}
  \label{fig1:ChromaSource}
\end{figure}

\medskip Therefore, the response of the baseline, or in a simpler way the voltage response of each antenna, is one of the most important aspect in the foreground avoidance method. It is crucial to control it, in order to limit the foreground leakage. Each additional reflection in the system time response spreads the foreground still further in the delay domain. Fig. \ref{fig1:ChromaSource} illustrates the main sources of chromaticity affecting the system and how they may impact the detection of the EoR delay power spectrum. Reflections can be divided into three categories: "internal" reflections between the elements of the antenna, reflections within the RF chain, and "external" reflections between the different antennas of the array. Internal reflections are mainly caused by standing waves which form inside the metal cage surrounding the feed, and between the dish and the backplane of the cage. Once received by the antenna, the signal propagates through the analogue receiver, which has its own response, and reflections occurring at the ends of coaxial cables are a common issue. The array is also very dense and the edges of the adjacent antennas are only 60 cm apart. Consequently, mutual coupling may be significant.

\medskip The differences between the expected delay power spectra obtained for the foreground and the EoR signal are illustrated in \citet{Ewall-Wice2016} and in \citet{Thyagarajan2016}, for various baselines. At very low delays and compared with the EoR signal, the foreground delay spectrum (cf. Equation \eqref{eq2:delSpectrum}) is four to five orders of magnitude higher, and its delay power spectrum (cf. Equation \eqref{eq3:delPoWSpectrum}) is about nine to ten orders of magnitude more intense. Thus, the goal is to attenuate the voltage time response of the system by a factor $10^{5}$, in order to limit the contamination of the EoR window. In the next section, we detail the model and parameters used to simulate the system response.

%

%%%%%%%%%%%%%%%%% SECTION 3 %%%%%%%%%%%%%%%%%%

\section{Parameters of the electromagnetic and electrical co-simulations} \label{sec:3.Co-simuPara}

%%%%%%%%%%%%%%%%% SECTION 3.1 %%%%%%%%%%%%%%%%%%

\subsection{Antenna model} \label{sec:3.1.Antenna}

\begin{figure}
  \centering
  \includegraphics[width=0.71\linewidth]{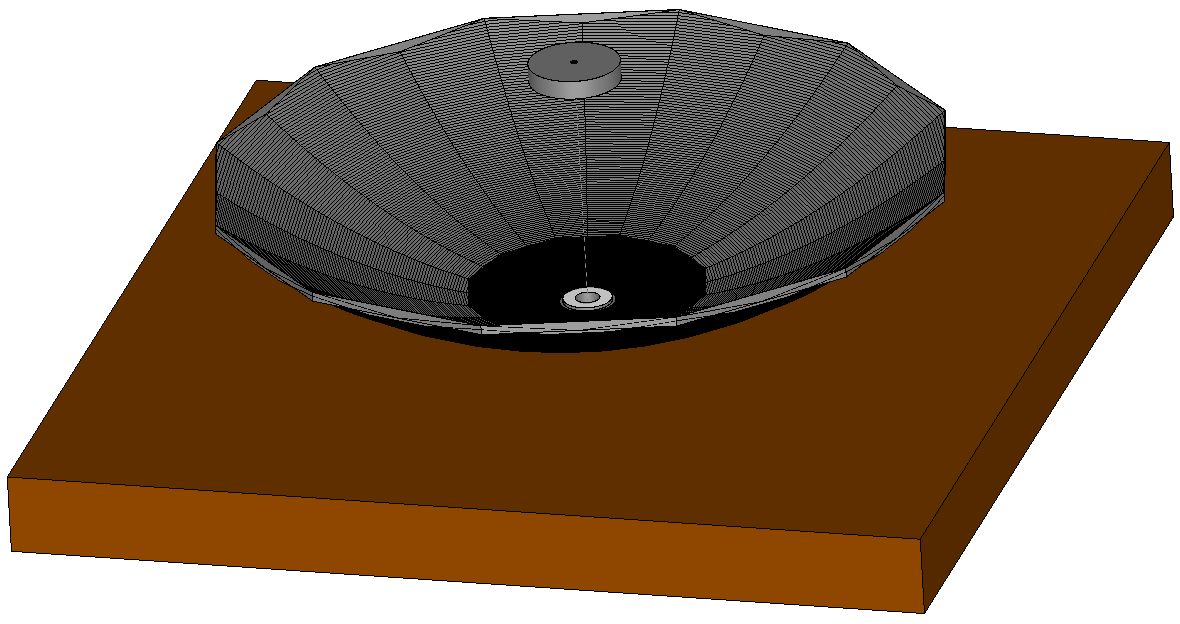}
  
  \vspace{1mm}
  
  \includegraphics[width=0.71\linewidth]{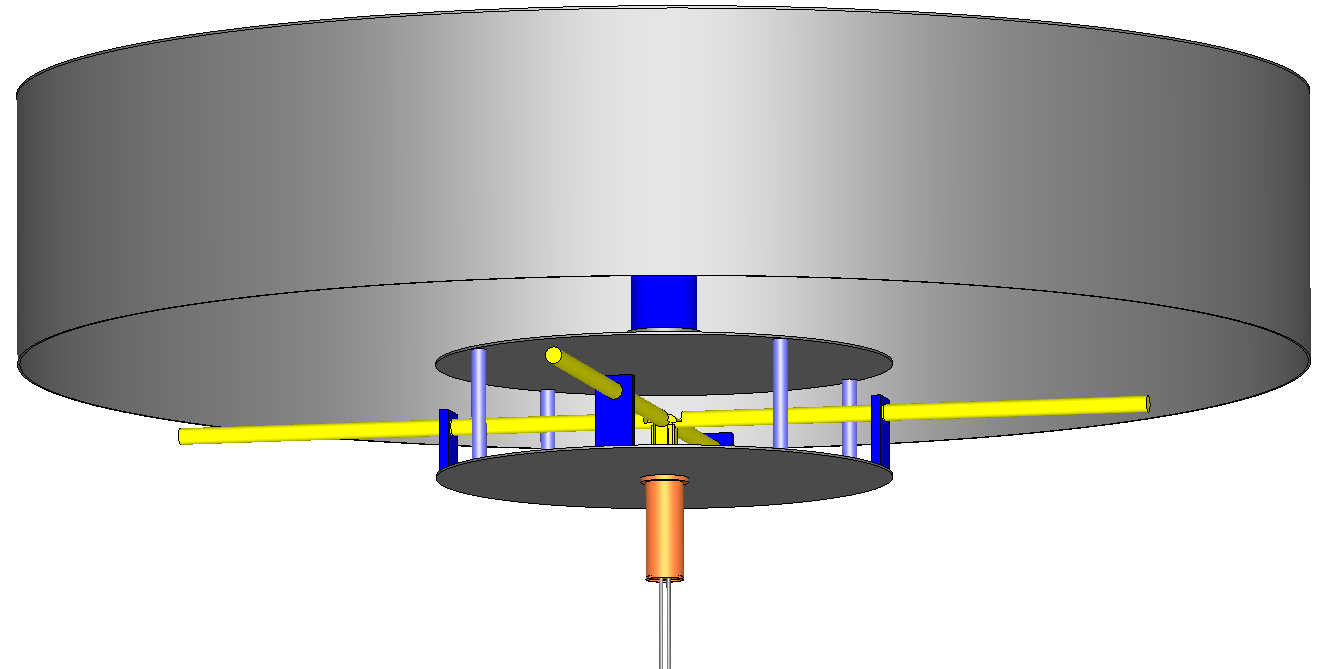}
  \caption{HERA dish and feed modelled with the simulation software CST.}
  \label{fig2:CSTmodel}
\end{figure}

\medskip Developed with CST (cf. Fig. \ref{fig2:CSTmodel}), the antenna model can be divided into 3 main elements: the reflector, the balanced feed, and a metal cage \citep{DeBoer2015, DeBoer2016}. The reflector is modelled by a faceted paraboloid, has a diameter of 14 m, and a focal ratio of 0.32. At its vertex, a concrete slab holds the PVC pipes supporting the structure of the antenna. In HERA Phase I, the dipole feed previously developed for PAPER is re-used. It consists of two perpendicular 1.3-m long dipoles surrounded by two aluminium discs, one below and one above, used to broaden the frequency response. The dipoles are terminated by four pins which are directly connected to the "front-end module" (FEM). Its presence is modelled by a brass tube terminated by two coaxial cables. The feed is surrounded by a cage which tapers the beam radiated by the dipoles. The cage is made up of two elements: a 172-cm backplane and a 36-cm high cylindrical "skirt". The feed is suspended 4.9 m above the dish. In our models, the metals have a finite conductivity, and the effects of the dielectric materials and the soil are also included.

%

%%%%%%%%%%%%%%%%% SECTION 3.2 %%%%%%%%%%%%%%%%%%

\subsection{Analogue receiver design} \label{sec:3.2.RFrec}

\medskip The receiver model is divided into three blocks: the FEM, the coaxial cable, and the "post-amplifier module" (PAM) (cf. Fig. \ref{fig1:ChromaSource}). A received electromagnetic wave is transformed into differential voltage signals by the dipole and the two 180\degr out-of-phase components are directly transmitted to the FEM. Inside this active balun, a transformer is used to adjust the FEM input impedance to the antenna impedance. The differential signals pass through three amplification stages, before being combined by a 180\degr hybrid coupler to form an unbalanced signal. The total gain is about 30 dB between 100 and 200 MHz. The unbalanced signals from the two polarizations are then transmitted to the PAM via a dual RG6U coaxial cable with an impedance of 75 $\Omega$. This 150-m cable attenuates the signals by 12.5 dB at 150 MHz. The PAM provides an additional gain of 66 dB. It consists of a transformer, five amplification stages, and a bandpass filter. In total, the receiver has a gain of 83.5 dB and a 3-dB bandwidth of 74 MHz centred on 148 MHz. The properties of this receiver developed by the NRAO are detailed in \citet{Parashare2006a, Parashare2006b, Parashare2007}. As for the noise parameters, the receiver noise is dominated by the first amplification stage which is made up of a NE461M02 transistor \citep{CEL}. The FEM has a minimum noise factor ${F_{\rm min}}$ which varies between 1.6 dB and 1.8 dB, an equivalent noise resistance ${R_{\rm N}}$ of about 27 $\Omega$, and an optimum source impedance ${Z_{\rm opt}}$ around 135 + 30j $\Omega$. The bandwidth of the receiver noise is also controlled by the bandpass filter.

%%%%%%%%%%%%%%%%% SECTION 3.3 %%%%%%%%%%%%%%%%%%

\subsection{Simulation parameters} \label{sec:3.3.SimuPara}

\medskip The properties of the antenna are simulated with the time domain solver of CST Microwave Studio which is based on the "finite integration technique" \citep{Weiland1977, Clemens2001}. This solver is efficient for problems with a complex geometry and for wideband simulations. An adaptive refinement method is used to control the meshing and validate that it does not affect the results. The antenna is described by a 2-port network to take into account the leakage between the two polarizations. A discrete termination port is used to excite the antenna with a broadband Gaussian pulse between 50 and 250 MHz. The simulation has a time step of 0.004 ns and is configured to stop after 1500 ns to characterize the voltage time response at high delays. We verify that the output voltage is attenuated by a factor $10^{5}$ at least. The differential S-parameters and the distance-independent electric farfield pattern $\bmath{E_{\rm pat}}\left( {\theta ,\;\varphi }\right)$ (in V) are extracted from the simulation. The beams are defined in spherical coordinates, and the farfield approximation is used to neglect the radial component. They are simulated every 0.5 MHz between 50 and 250 MHz, and with an angular step of 1\degr.

\medskip As for the RF receiver, the FEM, coaxial cable, and PAM are simulated with a microwave circuit simulation software, Genesys from Keysight Technologies \footnote{https://www.keysight.com}. The components are modelled using the library of Genesys and the data provided by the manufacturers. The goal of this circuit simulation is to have a flexible tool to estimate the S-parameters and the noise temperature of the receiver. These parameters are obtained when the amplifiers operate in linear mode; the effects of gain compression and saturation are not considered.

%%%%%%%%%%%%%%%%% SECTION 4 %%%%%%%%%%%%%%%%%%

\section{Co-simulation methodology} \label{sec:4.Methodology}

%%%%%%%%%%%%%%%%% SECTION 4.1 %%%%%%%%%%%%%%%%%

\subsection{Electrical circuit representation} \label{sec:4.1.EqCirc}

\begin{figure}
  \includegraphics[width=0.85\linewidth]{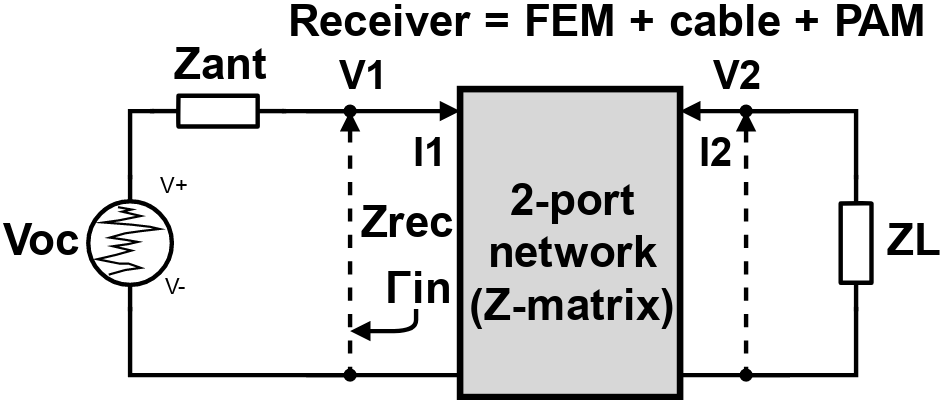}
  \caption{Equivalent electrical circuit of the antenna - RF receiver system.}
  \label{fig3:EquivCirc}
\end{figure}

\medskip The electrical properties of the antenna along with the receiver are described using the formalism used in RF network analysis. After receiving an electromagnetic wave, the antenna can be considered as a generator delivering a voltage ${V_{\rm oc}}$ with an internal complex impedance $Z_{\rm ant}$. The FEM, cable, and PAM are combined together to form a two-port network. The differential input is connected to the antenna, and the output to a 50-$\Omega$ load $Z_{\rm L}$. This system is equivalent to the electrical circuit in Fig. \ref{fig3:EquivCirc}. The network is characterized by the voltage ${V_{\rm 1}}$ and the current ${I_{\rm 1}}$ at the input port 1, and by ${V_{\rm 2}}$ and ${I_{\rm 2}}$ at the output port 2. By using the definition of the $Z$-parameters \citep[Ch. 4]{Pozar2011}, one can show that ${V_{\rm 2}}$ is equal to:
\begin{align}
    {V_{\rm 2}} = \frac{{{V_{\rm oc}}{Z_{\rm L}}{Z_{\rm 21}}}}{{({Z_{\rm L}} + {Z_{\rm 22}})\left( {{Z_{\rm ant}} + {Z_{\rm 11}}} \right) - {Z_{\rm 21}}{Z_{\rm 12}}}}.
	\label{eq6:Vout2}
\end{align}
The circuit can be further simplified by considering that the antenna "sees" an input impedance $Z_{\rm rec}$ which corresponds to the 2-port network along with the termination load. This equivalent impedance is used to describe the signal at the interface between the antenna output and the receiver input, while taking into account the transmission and reflection effects through the chain.
\begin{align}
    {Z_{\rm rec}} = {Z_{\rm 11}} - \frac{{{Z_{\rm 12}}{Z_{\rm 21}}}}{{{Z_{\rm L}} + {Z_{\rm 22}}}}.
	\label{eq7:Zrec}
\end{align}

%

%%%%%%%%%%%%%%%%% SECTION 4.2 %%%%%%%%%%%%%%%%%

\subsection{Validation of the models with measurements} \label{sec:4.2.Reflecto}

\begin{figure}
  \includegraphics[width=\linewidth]{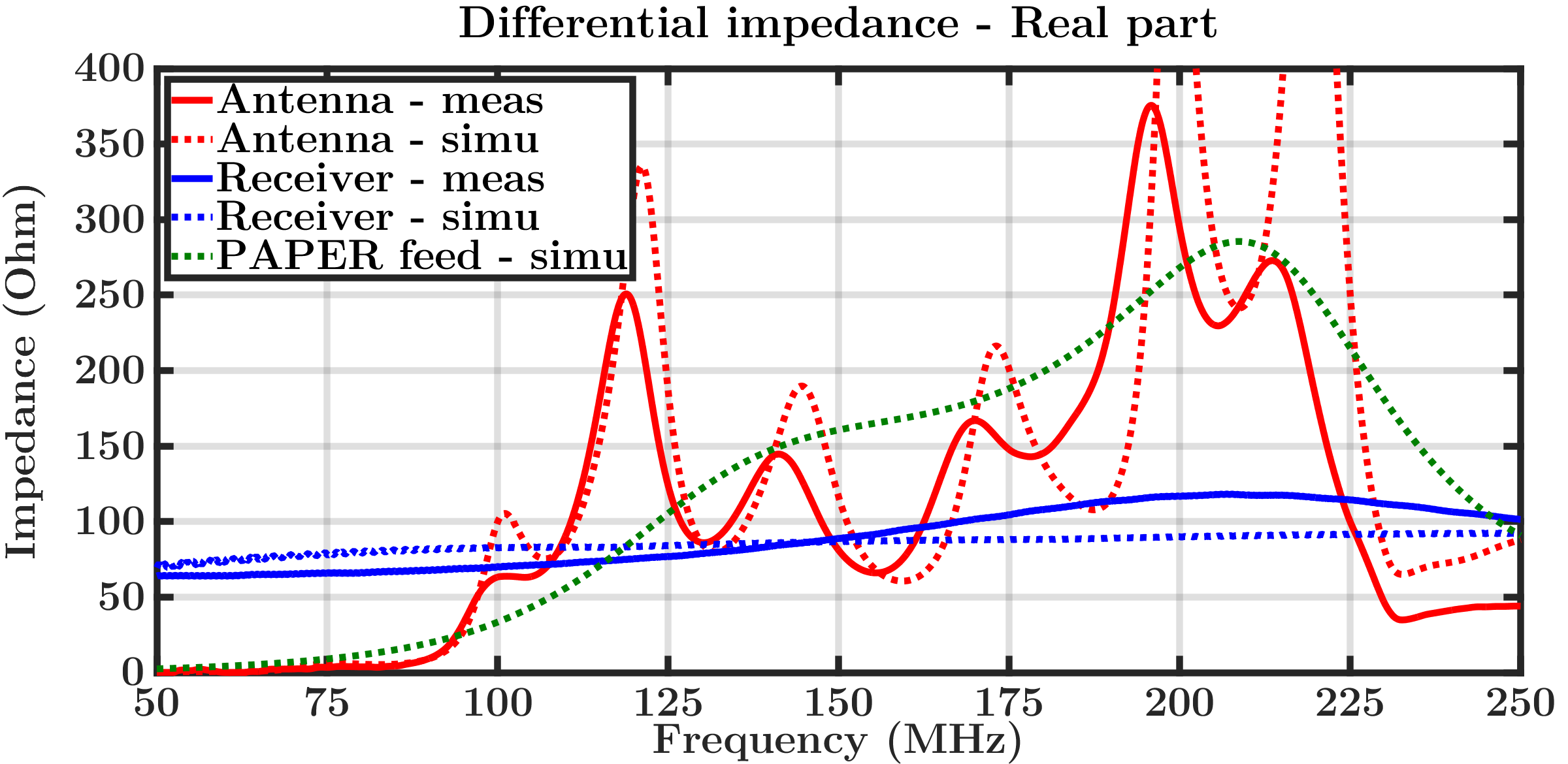}
  
  \vspace{3mm}
  
  \includegraphics[width=\linewidth]{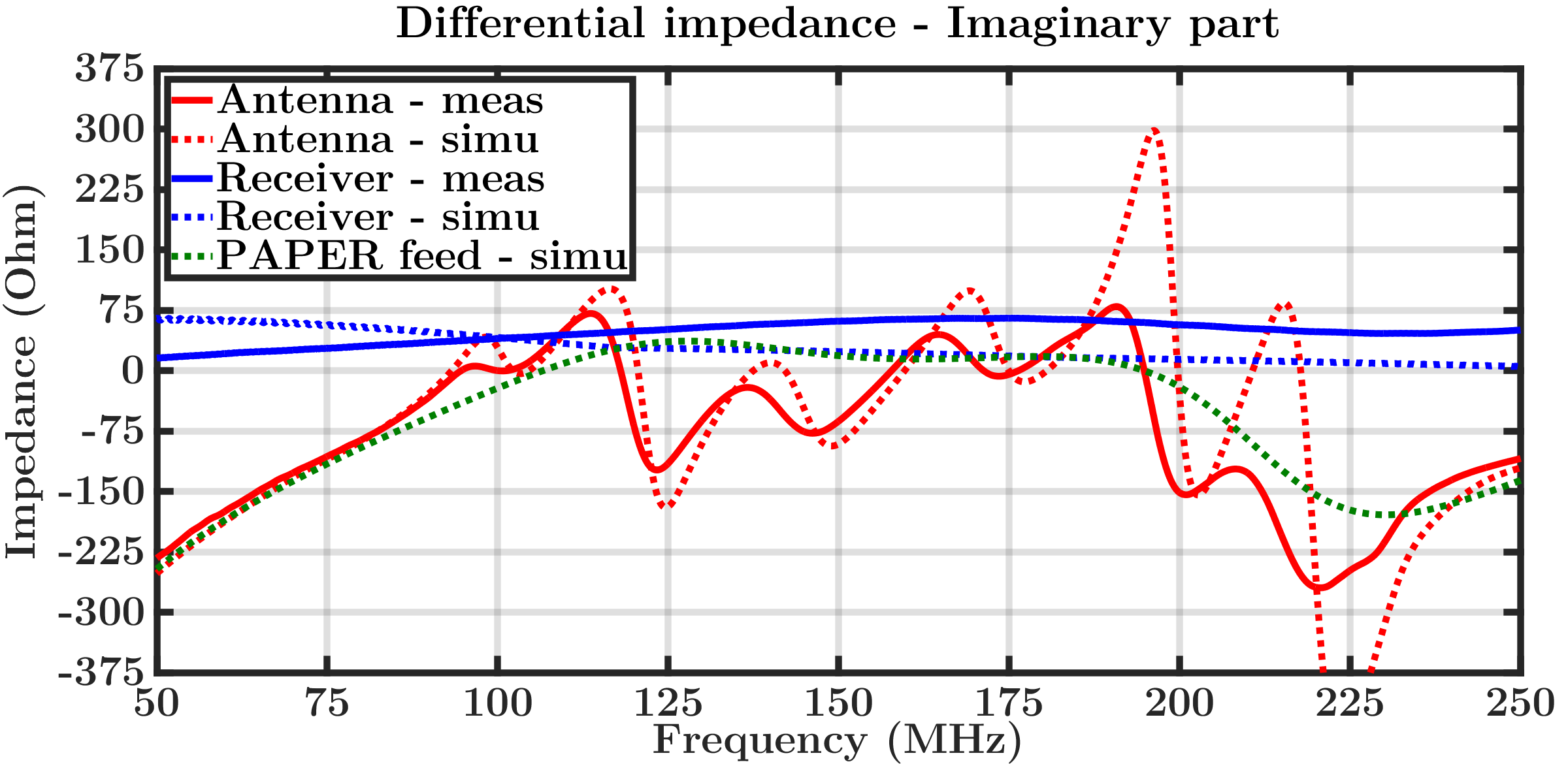}
  \caption{Differential input impedance of the RF receiver and of the antenna, from measurements ("meas") and simulations ("simu").}
  \label{fig4:antImp}
\end{figure}

\begin{figure}
  \includegraphics[width=\linewidth]{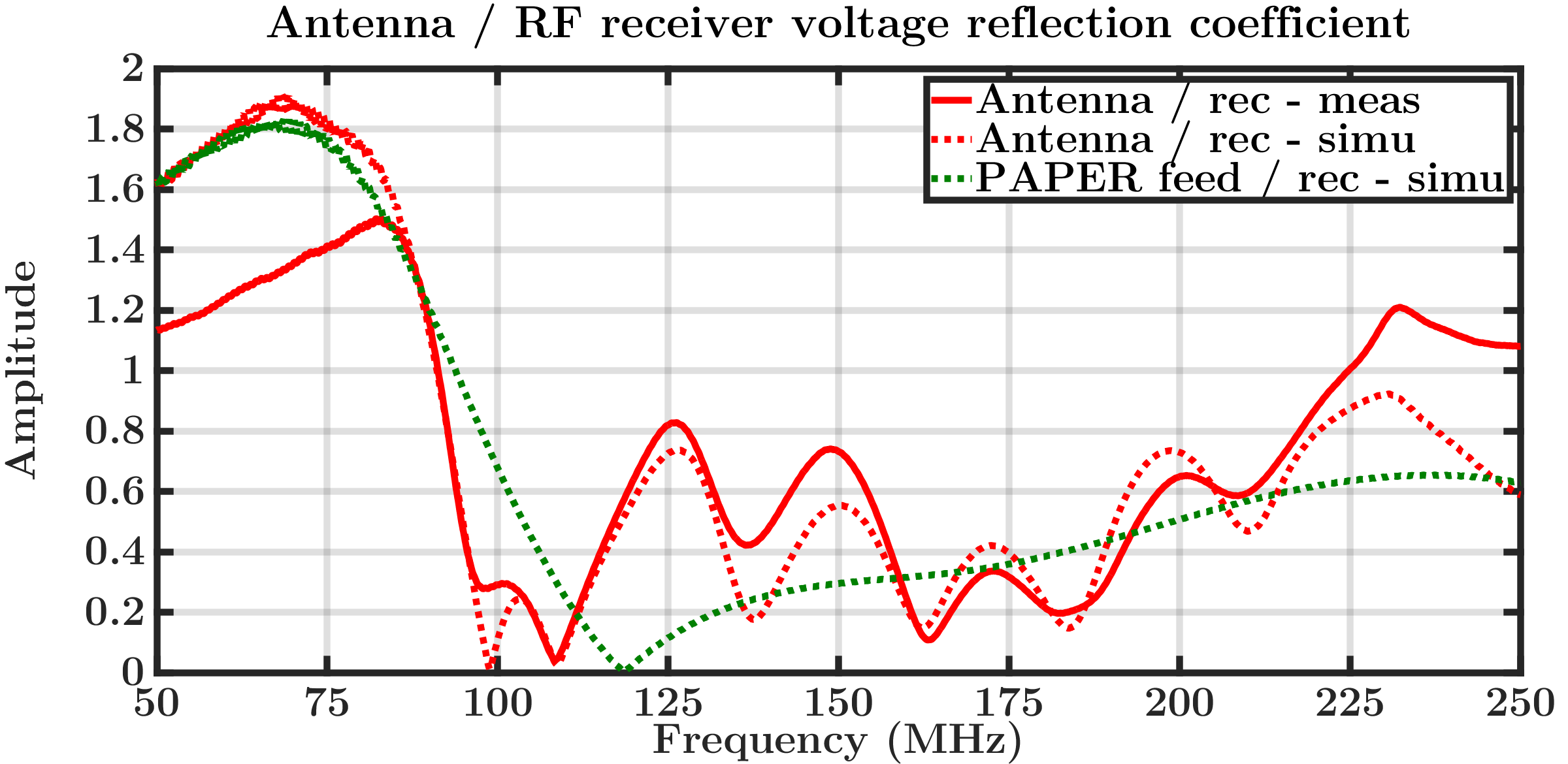}
  \caption{Amplitude of the voltage reflection coefficient at the interface between the antenna output and the receiver input.}
  \label{fig5:volRXcoefRec}
\end{figure}

\medskip A vector network analyzer (VNA) is used to measure the S and Z-parameters of each block of the receiver and of the antenna. The impedances obtained from measurements and simulations are presented in Fig. \ref{fig4:antImp}. The results are consistent, which is essential because we rely on numerical simulations to understand and improve the instrument. In the framework of the foreground avoidance method, reflectionless matching must be optimized in priority by avoiding major impedance discontinuities (real and imaginary parts) at the interface between two components \citep[Ch. 13]{Orfanidis2016}. The voltage reflection coefficient $\Gamma_{\rm in}$ at the antenna - receiver interface (cf. Fig. \ref{fig5:volRXcoefRec}) can be obtained as a function of their respective impedance thanks to Equation \eqref{eq8:VolRxCoef} \citep{Rahola2008}: 
\begin{align}
    \Gamma_{\rm in}  = \frac{{{Z_{\rm rec}} - {Z_{\rm ant}}}}{{{Z_{\rm rec}} + {Z_{\rm ant}}}}.
	\label{eq8:VolRxCoef}
\end{align}

\medskip One should not be surprised to see $\Gamma_{\rm in}$ with a magnitude superior to 1. This can occur with complex impedances and on condition that ${R_{\rm rec}}{R_{\rm ant}} + {X_{\rm rec}}{X_{\rm ant}} < 0$ \citep{Vernon1969}. However, the reflected power is always inferior to the incident power. The profile presents humps with a periodicity of about 30 MHz, which is associated with reflections occurring at 5 m, i.e. the distance between the cage and the vertex. By comparing this with the impedance of the PAPER feed alone, we can notice that the presence of the dish and cage has added sharp resonance peaks on the initial impedance. Such profile makes the impedance matching challenging and would require to insert a new matching network \citep{Fagnoni2016}. To complete Fig. \ref{fig5:volRXcoefRec}, let us add that $\Gamma$ is around 0.10 -- 0.15 at the FEM - cable interface as well as at the cable - PAM interface.

\medskip Lastly, \citet{Nunhokee2020} has measured the HERA radiation pattern in-situ using bright radio sources, and compared that with our simulated beams. The measured and simulated gains do agree well up to a level of -20 dB relative to the peak gain.

%

%%%%%%%%%%%%%%%%% SECTION 4.3 %%%%%%%%%%%%%%%%%

\subsection{System voltage response} \label{sec:4.3.SystVoltResp}

\medskip We now include the effects of the antenna beam pattern and its frequency response, in the description of the system. ${V_{\rm oc}}$ previously defined actually represents the "open-circuit voltage" produced by the antenna after being excited by an incident electric field $\bmath{{E_{\rm in}}} \left( {\theta ,\;\varphi } \right)$ when disconnected from the receiver. ${V_{\rm oc}}$ is obtained from the antenna effective length $\bmath{{l_{\rm eff}}} \left( {\theta ,\;\varphi } \right)$: 
\begin{align}
    {V_{\rm oc}}\left( {\theta ,\;\varphi } \right) = \bmath{{E_{\rm in}}} \left( {\theta ,\;\varphi } \right)\boldsymbol{\cdot}\bmath{{l_{\rm eff}}} \left( {\theta ,\;\varphi } \right).
	\label{eq9:Voc}
\end{align}
Thanks to the reciprocity principle, the effective length is identical in reception and in transmission. In the farfield approximation, it can be calculated with Equation \eqref{eq10:leff} \citep[Ch. 16]{Orfanidis2016}. 
\begin{align}
    \bmath{{l_{\rm eff}}} \left( {\theta ,\;\varphi } \right) = - \bmath{{E_{\rm pat}}} \left( {\theta ,\;\varphi } \right)\frac{{4\pi }}{{k{Z_{\rm fs}}{I_{\rm p}}}}\mathrm{j}\;,
	\label{eq10:leff}
\end{align}
with $\bmath{{E_{\rm pat}}}\left( {\theta ,\;\varphi } \right)$ the electric farfield pattern simulated with CST, ${Z_{\rm fs}}$ the impedance of free space equal to $\sqrt{{\mu_{\rm 0}}/{\varepsilon_{\rm 0}}} \approx \rm{377 \;\Omega}$, and ${I_{\rm p}}$ the current at the feeding point of the dipole. In simulation, if the antenna is excited by an input travelling wave ${a_{\rm in}}$ coming from a differential port with an impedance ${Z_{\rm 0}}=100 \;\Omega$, then ${I_{\rm p}}$ is equal to: 
\begin{align}
    {I_{\rm p}} = \frac{{2{a_{\rm in}}\sqrt {{Z_{\rm 0}}} }}{{{Z_{\rm ant}} + {Z_{\rm 0}}}}.
	\label{eq11:Iport}
\end{align}

\medskip Similarly to Section \ref{sec:2.1.DelSpec}, we define the system voltage response in reception $\mathbfit{H}\left( {\theta ,\;\varphi } \right)$ which links the receiver voltage output ${V_{\rm 2}}\left( {\theta ,\;\varphi } \right)$ to the incident electric field $\bmath{{E_{\rm in}}} \left( {\theta ,\;\varphi } \right)$.
\begin{align}
    {V_{\rm 2}}\left( {\theta ,\;\varphi } \right) = \mathbfit{H}\left( {\theta ,\;\varphi } \right)\boldsymbol{\cdot}\bmath{{E_{\rm in}}} \left( {\theta ,\;\varphi } \right).
	\label{eq12:transFunc}
\end{align}
Thanks to Equation \eqref{eq6:Vout2} and the antenna effective length, ${V_{\rm 2}}\left( {\theta ,\;\varphi } \right)$ can be calculated for any incidence angle and polarization. Thus, the expression of the system voltage response in reception and in the frequency domain can be derived as a function of the antenna impedance, its E-farfield pattern, and the receiver 2x2 impedance matrix $\mathbf{Z_r}$. Lastly, by Fourier transforming the obtained Equation \eqref{eq13:VolSystResp}, the direction-dependent chromatic effects can be quantified in the time domain. The results are presented in Section \ref{sec:5.RecCplPerf}.
\begin{align}
    \mathbfit{H}\left( {\theta ,\;\varphi } \right) &=  - \bmath{{E_{\rm pat}}} \left( {\theta ,\varphi } \right)\left[ {\frac{{4\pi }}{{k{Z_{\rm fs}}}}\mathrm{j}} \right] \nonumber \\
    &\left[ {\frac{{{Z_{\rm ant}} + {Z_{\rm 0}}}}{{2{a_{\rm in}}\sqrt {{Z_{\rm 0}}} }}} \right]\left[ {\frac{{{Z_{\rm L}}{Z_{\rm r21}}}}{{\left( {{Z_{\rm L}} + {Z_{\rm r22}}} \right)\left( {{Z_{\rm ant}} + {Z_{\rm r11}}} \right) - {Z_{\rm r21}}{Z_{\rm r12}}}}} \right].
	\label{eq13:VolSystResp}
\end{align}
%

%%%%%%%%%%%%%%%%% SECTION 4.4 %%%%%%%%%%%%%%%%%

\subsection{Mutual coupling simulations} \label{sec:4.4.MutualCoup}

\begin{figure}
  \includegraphics[width=\linewidth]{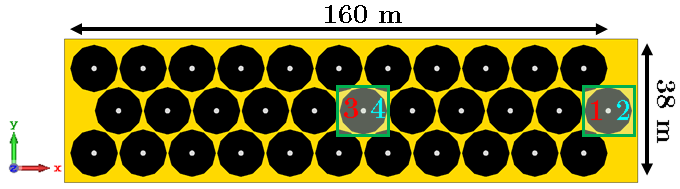}
  \caption{Array configuration used to simulate the mutual coupling in HERA with CST. The red numbers correspond to the ports associated with the X-polarization, and the blue numbers with the Y-polarization.}
  \label{fig6:confArray}
\end{figure}

\medskip The adjacent antennas are only 60 cm apart in the array. Therefore, mutual coupling plays an important role in the chromaticity of the telescope \citep{Craeye2011}. The incident signal is scattered by the structure of the dishes, and the signal absorbed by a dipole may also be re-radiated, because of impedance mismatch. The system response associated with an antenna is analysed for the array configuration presented in Fig. \ref{fig6:confArray}. This configuration was chosen because preliminary studies showed that the coupling between two antennas is the most intense when their dipoles are parallel and aligned. In terms of S-parameters, the level of coupling between two dipoles directly parallel is about -45 dB. Moreover, the power exchanged between orthogonal dipoles, i.e. between the X and Y-polarisations of two antennas, is also significant when two adjacent antennas form an angle of 60\degr with the X-axis. In this case, the level of coupling is around -50 dB. The coupling effects on the Y-polarization are studied in this paper, but the conclusions are also valid for the X-polarisation. The Y-polarization of an antenna is mainly coupled with the other dipoles aligned in the X-direction and with the rows directly above and below. Thus, the array is expanded along the X-axis and includes 11 columns. This configuration corresponds to half of the final length of the core, and the size is actually limited by the available computational resources (250 GB of RAM).

\medskip The S-parameters provide information about mutual coupling, but in transmission. The next step is to describe the system in reception with N antennas \citep{Lui2009}. To compute the response associated with a receiver, this 2N-port network can be simplified into a 2-port network similarly to Fig. \ref{fig3:EquivCirc}. The reference receiver still sees an input impedance ${Z_{\rm ant}}$ and a termination impedance of 50 $\Omega$. However, ${Z_{\rm ant}}$ needs to be modified to combine the effects of the connected antenna and the rest of the array. We assume that each antenna port is terminated by the same impedance ${Z_{\rm term}}$. The voltage and the current at each port of this 2N-port system are related by the $Z$-matrix, and the impedance ${Z_{\rm ant}}$ seen by the reference port $m$ can be obtained by solving the following system of equations:
\begin{align}
    {V_{\rm 1}} = {Z_{\rm 11}}{I_{\rm 1}} +  \ldots  + {Z_{\rm 1m}}{I_{\rm m}} +&  \ldots  + {Z_{\rm 12N}}{I_{\rm 2N}} =  - {Z_{\rm term}}{I_{\rm 1}} \nonumber \\
	&\vdots \nonumber \\
	{V_{\rm m}} = {Z_{\rm 1m}}{I_{\rm 1}} +  \ldots  + {Z_{\rm mm}}{I_{\rm m}} +&  \ldots  + {Z_{\rm 12N}}{I_{\rm 2N}} = {Z_{\rm ant}}{I_{\rm m}} \nonumber \\
	&\vdots \nonumber \\
	{V_{\rm 2N}} = {Z_{\rm 2N1}}{I_{\rm 1}} +  \ldots  + {Z_{\rm 2Nm}}{I_{\rm m}} +&  \ldots  + {Z_{\rm 2N2N}}{I_{\rm 2N}} =  - {Z_{\rm term}}{I_{\rm 2N}}.
	\label{eq14:ZantCoup}
\end{align}
In practice, ${Z_{\rm ant}}$ in Equation \eqref{eq11:Iport} is defined with respect to the termination impedance used in CST to generate the beams with coupling, i.e. $Z_{\rm term}=Z_{\rm 0}=100 \;\Omega$. Then, in order to include the effects of the other receivers, ${Z_{\rm ant}}$ in Equation \eqref{eq6:Vout2} needs to be recalculated using the termination impedance ${Z_{\rm rec}}$. Thus, Equation \eqref{eq13:VolSystResp} is still valid, but on condition that ${Z_{\rm ant}}$ is properly adjusted. 

%

%%%%%%%%%%%%%%%%% SECTION 5 %%%%%%%%%%%%%%%%%%%

\section{Effects of the RF receiver and mutual coupling on the system performance} \label{sec:5.RecCplPerf}

%%%%%%%%%%%%%%%%% SECTION 5.1 %%%%%%%%%%%%%%%%%

\subsection{Mismatch and system noise temperature} \label{sec:5.1.NoiseSensi}

\begin{figure}
  \includegraphics[width=1\linewidth]{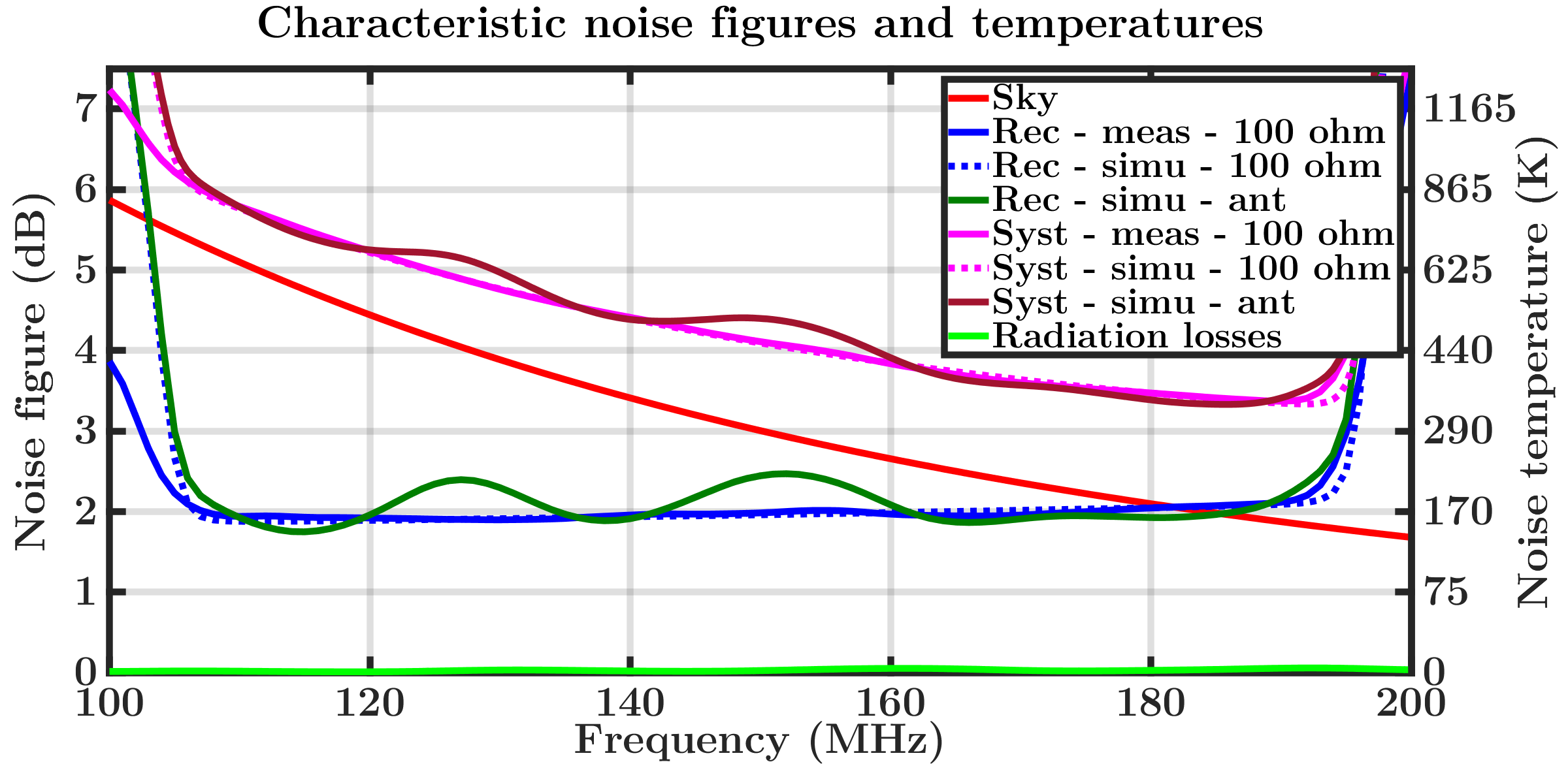}
  \caption{Noise generated by the sky, the radiation losses, and the simulated / measured receivers when terminated by 100 $\Omega$ or the measured antenna impedance. Combined together, they define the system noise temperature.}
  \label{fig7:characTemp}
\end{figure}

\medskip In addition to the reflection properties of the receiver, it is also important to consider the noise it generates, since this directly affects the antenna sensitivity. The noise generated by an amplifier varies as a function of its source impedance \citep[Ch. 12]{Pozar2011}. Therefore, the receiver has to be connected to the antenna impedance in order to obtain accurate noise data. In practice, it is difficult to achieve with direct noise measurements. For this reason, we rely on our co-simulation to include the effects of the mismatch between the optimum noise impedance of the receiver and the antenna impedance. This approach is then validated by comparing the noise temperature obtained from measurements ("meas") and simulations ("simu"), when the receiver is connected to 100 $\Omega$, the differential impedance of the measurement device. The effects of the variations in the input impedance are illustrated in Fig. \ref{fig7:characTemp}. With respect to 100 $\Omega$, the simulated and measured receiver noise temperatures are stable, around 160 K, and do agree well. On the other hand, with the antenna impedance, the noise temperature varies up to 220 K at 125 MHz and 150 MHz. Note that below 105 MHz and above 195 MHz, the receiver temperature is controlled by the passband filter. 

\medskip In this paper, we assume that the system noise temperature ${T_{\rm sys}}$ is equal to the receiver temperature ${T_{\rm rec}}$, plus the noise temperature effectively received from the sky, plus a term associated with the losses caused by the dissipative elements of the antenna at the ambient temperature ${T_{\rm 0}}$ = 290 K and with the radiation efficiency $\eta _{\rm rad}$ = 99\% \citep{Cortes-Medellin2007}:
\begin{align}
    {T_{\rm sys}} = {T_{\rm rec}+{\eta _{\rm rad}}T_{\rm sky}+\left(1-{\eta _{\rm rad}}\right)T_{\rm 0}}.
	\label{eq15:Tsys}
\end{align}
To simplify the study, we also consider that the sky contribution is uniform. For a cold patch of sky typically observed by HERA far from the galactic centre and at low frequencies, the sky temperature ${T_{\rm sky}}$ can be approximated by a power law \citep{Furlanetto2015}:
\begin{align}
    {T_{\rm sky}} = 180{\left( {\frac{\upsilon }{{180\;\rm{MHz}}}} \right)^{-2.6}}.
	\label{eq16:Tsky}
\end{align}

\medskip Fig. \ref{fig7:characTemp} shows that the system temperature is dominated by the sky contribution at low frequency. However, as the frequency increases, the sky temperature decreases and the effects of the receiver become preponderant above 180 MHz. For the purpose of comparison, the system noise temperature is also calculated using the receiver temperatures obtained with the antenna impedance and with 100 $\Omega$. In particular, it is about 10\% higher at 125 and 150 MHz with the antenna impedance. Thus, with respect to the 100-$\Omega$ reference impedance, this would require to extend the integration time of the data by about 20\% in order to keep the same rms error.

%%%%%%%%%%%%%%%%% SECTION 5.2 %%%%%%%%%%%%%%%%%

\subsection{Effects of the receiver on the system response} \label{sec:5.2.CableSystResp}

\begin{figure}
  \includegraphics[width=\linewidth]{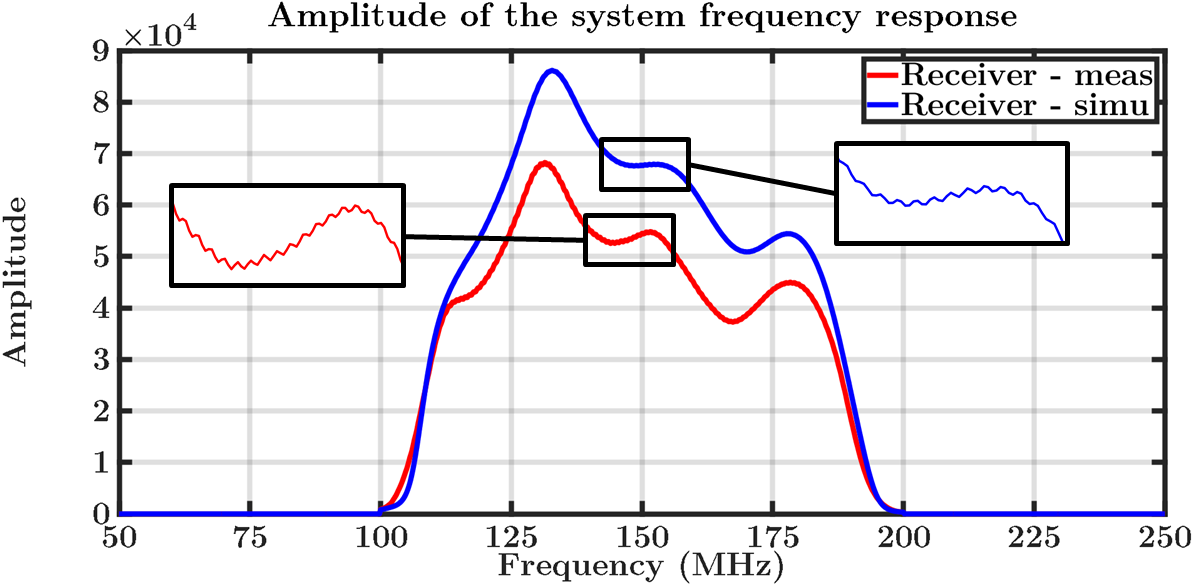}
  \caption{Amplitude of the frequency voltage response of the system $H(f)$ at zenith, with the measured and simulated RF receivers.}
  \label{fig8:systRespFreq}
\end{figure}

\medskip In this section, we study the effects of the receiver on the system voltage response. The effects of mutual coupling are added in the next section, in order to better understand their individual impact on the delay power spectrum. Equation \eqref{eq13:VolSystResp} is used to calculated the frequency response to an incident plane wave coming from the zenith. Fig. \ref{fig8:systRespFreq} presents the results with the measured and simulated receivers. By zooming in on the frequency spectra, small and fast ripples are visible. This is typical in mismatched transmission lines, and is caused by destructive and constructive interferences. These ripples have a periodicity of about 0.8 MHz, which corresponds to reflections occurring in a 150-m cable with a velocity factor of 0.82. 

\begin{figure}
  \includegraphics[width=0.99\linewidth]{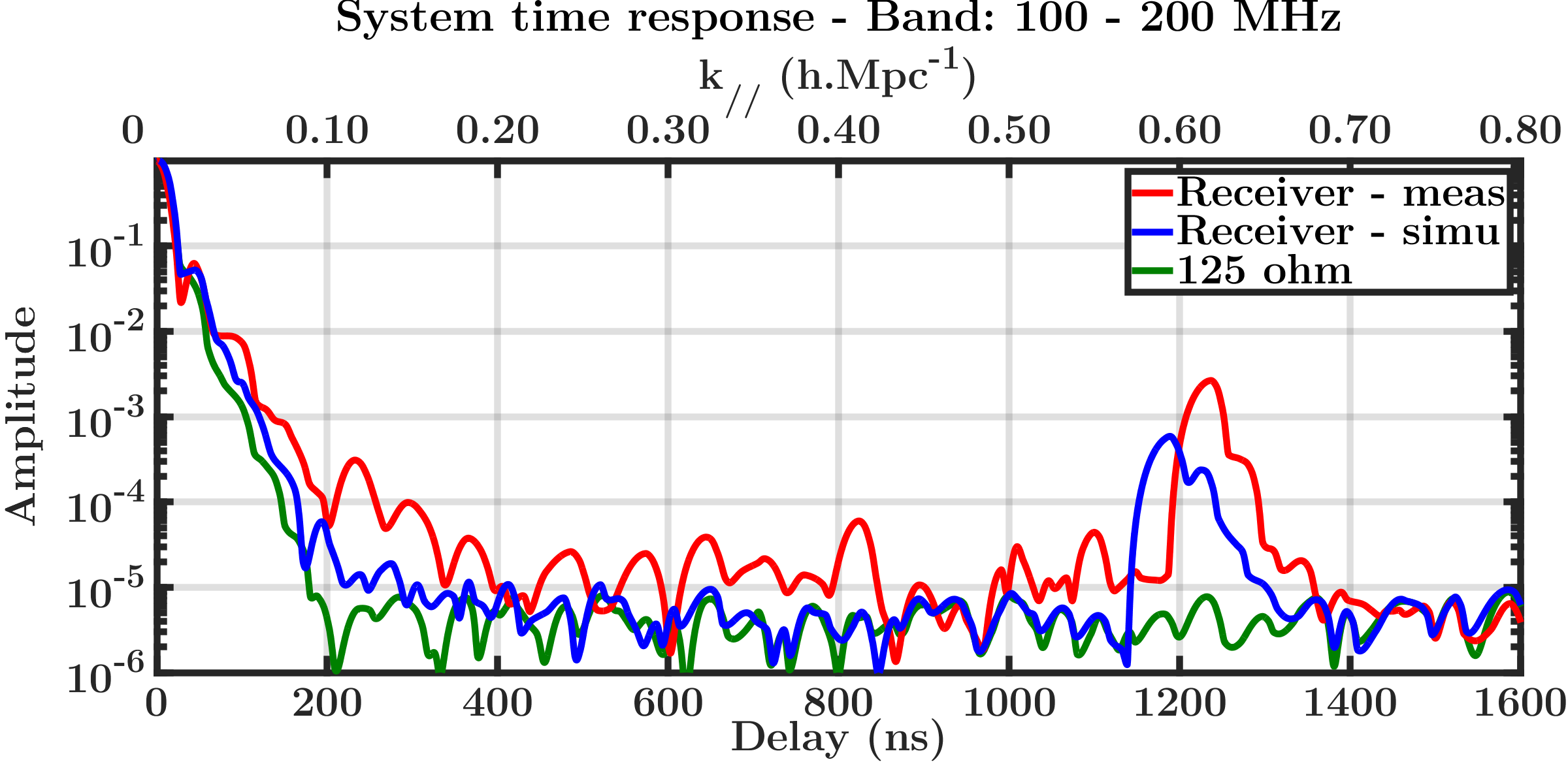}
  \caption{Time voltage response of the system $h(\tau)$ at zenith, normalized and windowed by a Blackman-Harris function between 100-200 MHz.}
  \label{fig9:systResTimeZen}
\end{figure}

\medskip Before being Fourier transformed, the spectrum is windowed by a Blackman-Harris function to mitigate the effects of the leakage associated with the 100 -- 200 MHz bandwidth of the correlator. Thus, the noise floor of the voltage time response presented in Fig. \ref{fig9:systResTimeZen} is below $10^{-5}$. The response is computed when the antenna is terminated by a 125-$\Omega$ load as in \citet{Ewall-Wice2016}, and by the transmission parameters of the receiver. When terminated by 125 $\Omega$, the system (cf. Fig. \ref{fig3:EquivCirc}) is equivalent to a generator $V_{\rm oc}$ in series with $Z_{\rm ant}$ and $Z_{\rm L}$ = 125 ohm, and its response is defined with respect to the voltage at $Z_{\rm L}$. The response is also normalized.

\medskip At low delays, the signal is reflected every 30 ns; this delay is associated with the reflections between the cage and the vertex of the dish. The response is quickly attenuated by a factor $10^{4}$, and below about 250 ns, the attenuation mainly depends on the quality of the impedance matching between the antenna and the FEM. However, at higher delays the response of the measured receiver remains about 5 to 10 times more intense. This is due to the chromaticity of the cable. Unlike an ideal cable used in the simulated receiver, small variations in its physical properties, for instance when it has been bent, locally modify its characteristic impedance, which causes micro-reflections \citep{Zhu2019}. This illustrates the importance of having high quality cables and as straight as possible. Lastly, because of the reflection of the signal at the end of the 150-m cable, the response increases by a factor 300 between 1200 -- 1300 ns.

\begin{figure}
  \includegraphics[width=\linewidth]{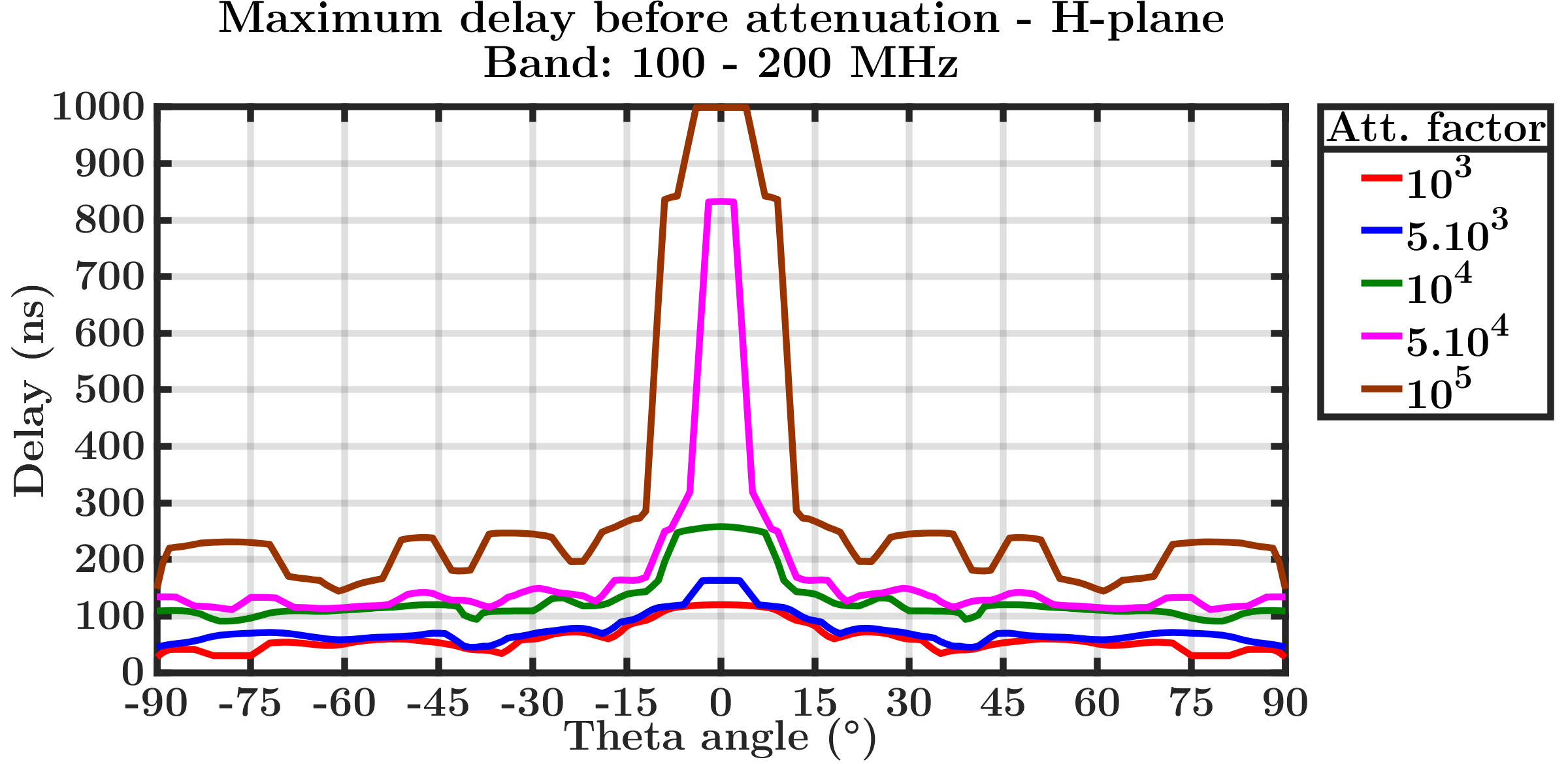}
  \caption{Maximum delays before the system response is attenuated by a certain factor, with the measured receiver, and for the H-plane.}
  \label{fig10:maxDel}
\end{figure}

\medskip The attenuation of the system time response as a function of the incidence angle is then studied. The goal is to transpose the spatial and frequency variations of the beam into the delay domain, in order to quantify the contribution from the sidelobes. The results obtained in the H-plane with the measured receiver are presented, since the sidelobes are more intense in this plane (cf. Fig. \ref{fig11:chromaBeamCoupl3D}). Fig. \ref{fig10:maxDel} represents the maximum delay after which the system time response is attenuated by a given factor. The thresholds are calculated over the first 1000 ns. Thus, the signal reflection at the end of the cable is disregarded, and it is assumed that it will be possible to calibrate out this effect \citep{Kern2020b}. By considering excitation signals with similar spectra, the response generated by the sidelobes is about 10 times lower than the response at zenith at high delays. In other words, for a single antenna if we compare two different sources, one at zenith and the other one in a sidelobe, the response associated with the sidelobe is more chromatic if the intensity of its source is at least one order of magnitude higher than the source at zenith.

%

%%%%%%%%%%%%%%%%% SECTION 5.3 %%%%%%%%%%%%%%%%%%

\subsection{System response including mutual coupling} \label{sec:5.3.CoupSystResp}

\begin{figure}
    \centering
    \includegraphics[width=1\linewidth]{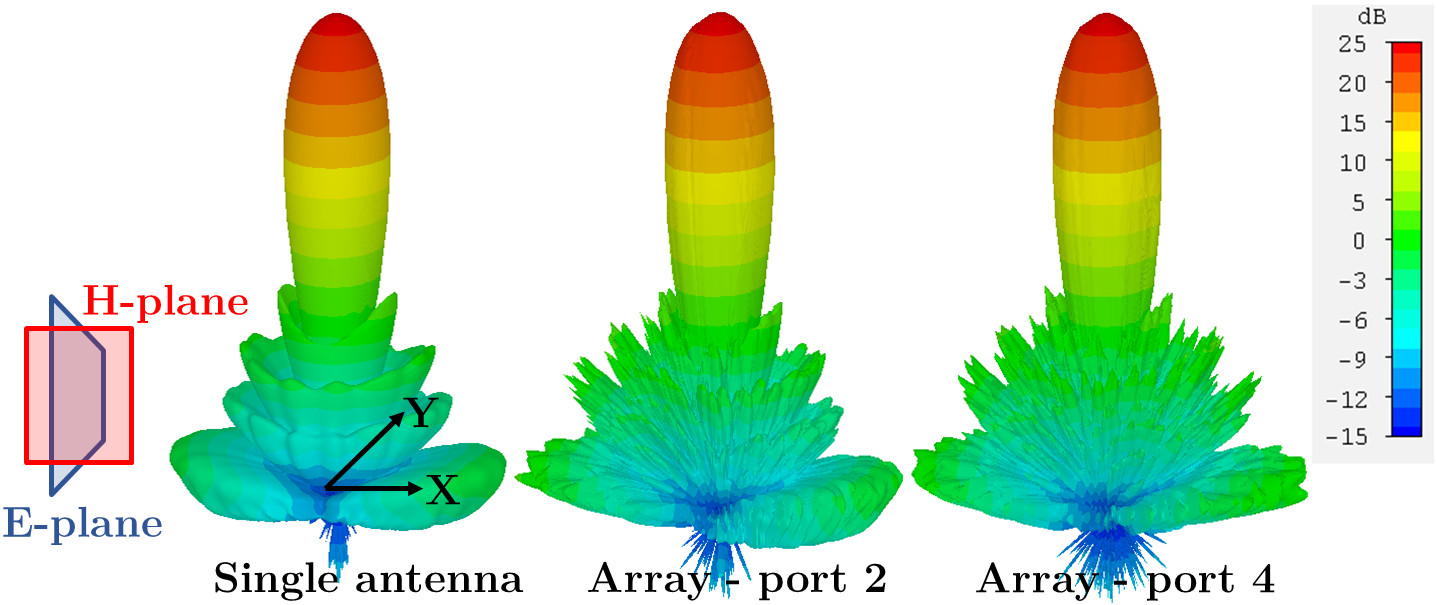}
    \caption{Gain patterns at 150 MHz (amplitude in dB) without and with mutual coupling in the simulated strip configuration, Y-polarisation.}
    \label{fig11:chromaBeamCoupl3D}
\end{figure}

\begin{figure}
  \includegraphics[width=\linewidth]{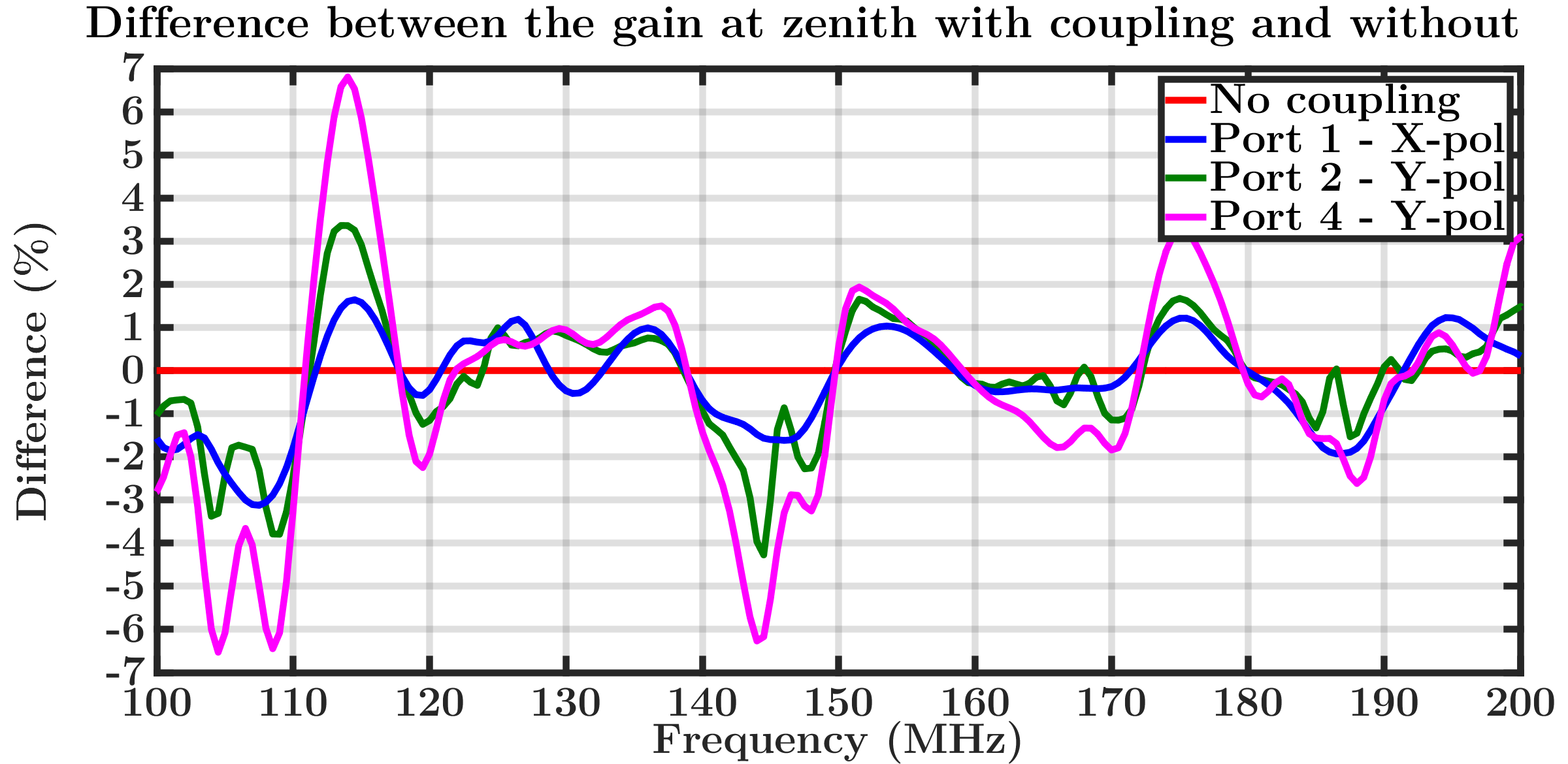}
  \caption{Percentage of difference in the simulated gain at zenith with respect to a system without mutual coupling.}
  \label{fig12:gainDiff}
\end{figure}

\medskip In a dense array, mutual coupling significantly affects the beam radiated by the antennas, depending on their position. To illustrate that, Fig. \ref{fig11:chromaBeamCoupl3D} shows how the gain pattern of the reference antennas at the centre and at the edge of the studied array (cf. Fig. \ref{fig6:confArray}) is modified at 150 MHz. The adjacent antennas are responsible for additional sidelobes and ripples on the radiation pattern. On average, without mutual coupling the maximum sidelobe level is about -23 dB, and with mutual coupling this level is increased by 2 to 4 dB. Moreover, each antenna "sees" a slightly different electromagnetic environment, which causes disparities between the beams, and in particular in the main lobe. Depending on the antenna position, the 3-dB beamwidth varies up to 1\degr, and the gain at zenith fluctuates by $\pm$ 0.3 dB, which represents a difference of up to 7\% (cf. Fig. \ref{fig12:gainDiff}). The beam becomes asymmetric for the antennas at the edges of the array. In the case of redundant-baseline calibration, all these elements matter. \citet{Orosz2018} simulated the impact of antenna-to-antenna variations on the estimation of the delay power spectrum by modelling the beam with an Airy function, and showed that such differences can have prejudicial effects. Consequently, the antennas with the largest deviations in their response may have to be excluded from the calibration process, especially the ones at the edges. 


\begin{figure}
  \includegraphics[width=\linewidth]{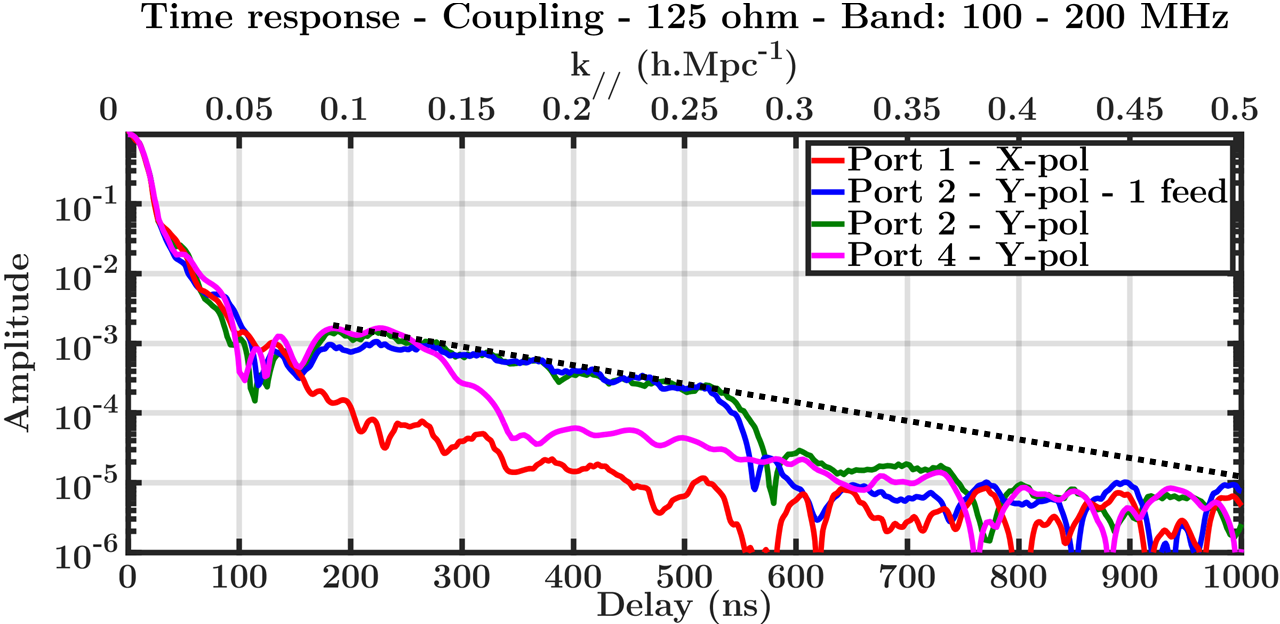}
  \caption{System time response at zenith, terminated by a 125-$\Omega$ load and including mutual coupling for various antenna configurations.}
  \label{fig13:systRespCouplNoRec}
\end{figure}

\medskip In order to only quantify the impact of the mutual coupling on the system time response, the antennas are terminated by a 125-$\Omega$ load. Fig. \ref{fig13:systRespCouplNoRec} presents the results for the antennas at the centre and at the edge of the array model. This plot shows that the mutual coupling significantly affects the system performance and is actually the dominant effect once the response is attenuated by a factor $10^{3}$. Then, the signal slowly decays and eventually drops after a delay which corresponds to the maximal distance between the reference antenna and the edges of the simulated array. Thus, the attenuation of the response depends on the antenna position. When the mutual coupling dominates the response, reflections with a periodicity of about 50 ns can be observed, which corresponds to the distance between two dish centres, i.e. 14.6 m. Every 50 ns, the contribution from one of the aligned dishes is received by the reference antenna. If the simulated antenna strip were longer, the extrapolation of the slope of the system response suggests that it would be attenuated by a factor $10^{4}$ between 600 and 700 ns, and by a factor $10^{5}$ after about 1000 ns. By comparison, the core of the final array is about 300-m large along the X-axis and 250-m large along the Y-axis. In terms of propagation delay, this is respectively equivalent to 1000 ns and 833 ns. Therefore, this study indicates that in the final array any antenna could significantly interact with the dishes at the edges. In the worst case, the response of an antenna at the edge will not be attenuated by a factor $10^{5}$ before 1000 ns. In the best case for the central antenna, its response will start dropping below $10^{-4}$ only after about 500 ns. Lastly, the array is simulated with only one feed at its edge. In this configuration, the time response of the port 2 with one feed (blue curve) is comparable to the response when all the feeds are present (green curve). Thus, the mutual coupling in reception is mainly caused by feed-dish interactions, rather than by feed-feed interactions.

\begin{figure}
  \includegraphics[width=\linewidth]{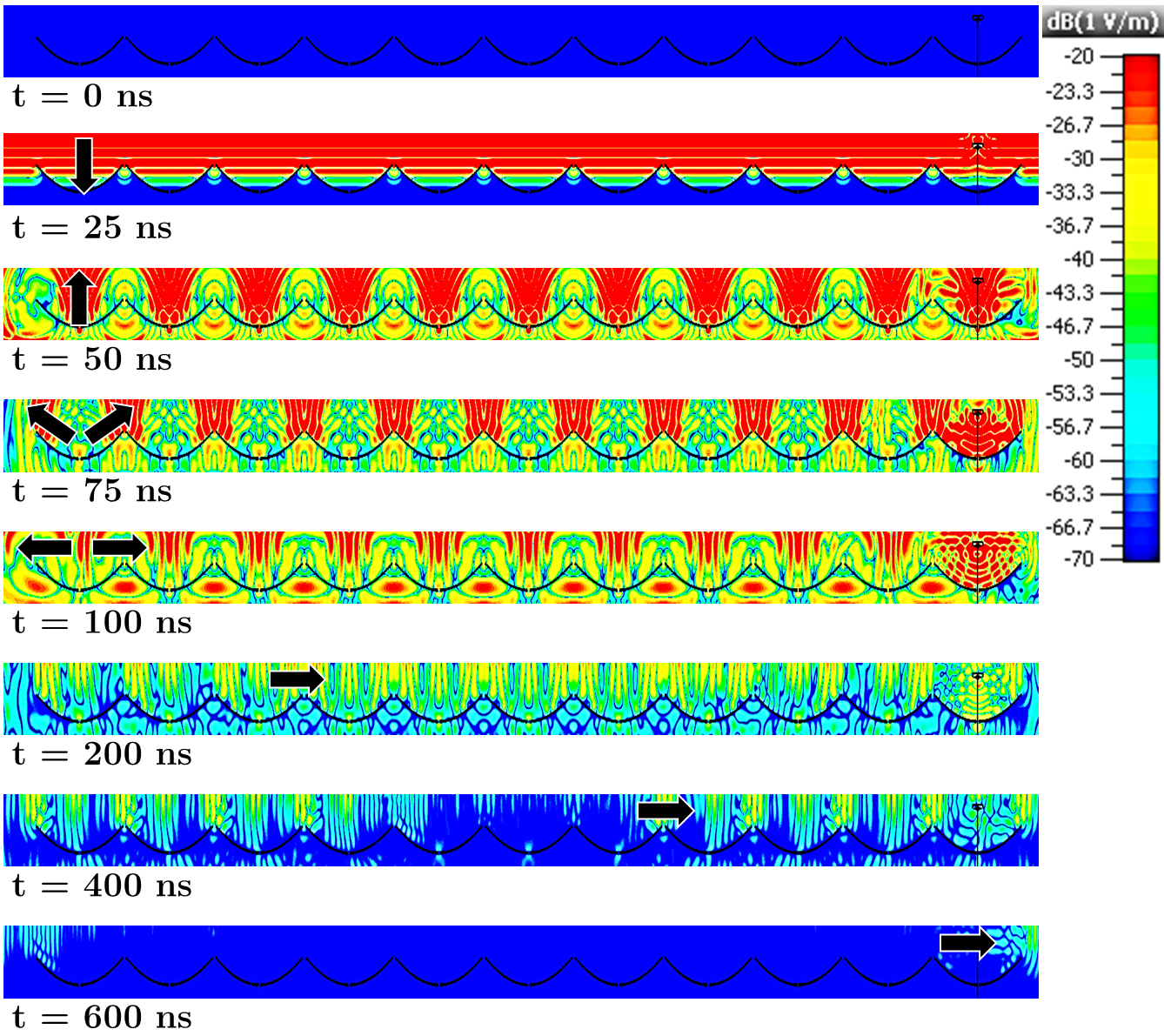}
  \caption{Snapshots of the electric field propagating through the array, when the antennas are excited by a plane wave coming from the zenith, at t = 0 ns, 25 ns, 50 ns, 75 ns, 100 ns, 200 ns, 400 ns, and 600 ns. The intensity of the E-field is expressed in dBV/m. Note how a part of the incident electromagnetic wave is reflected upwards and the other part towards the edge of the dishes. Then, the E-field propagates from one dish to the next one up to the end of the array, and with little attenuation.}
  \label{fig14:E-fieldTimeCoupl}
\end{figure}

\medskip This analysis is confirmed by the study of the electric field propagating through the antenna array. Fig. \ref{fig14:E-fieldTimeCoupl} presents several snapshots obtained with CST when the array is excited by a plane wave coming from the zenith and with a Y-polarization. All the feeds have been removed except one at the far right, in order to analyse the propagation mechanisms between the dishes. The incident wave reaches the antennas at t = 25 ns, before being reflected by the dishes at about t = 50 ns towards the zenith. However, a significant fraction of the wave is also reflected at 90\degr towards the edges of the dishes, as we can see at t = 75 ns. Then, the reflected waves are able to propagate from one dish to another one, with little attenuation. Eventually, the waves coming from the antennas at the far left (or far right) reach the opposite antennas after 500 -- 600 ns. The central dishes are less affected since the propagation delay between these antennas and those at the edges is shorter. The amplitude of the incident electric field is 1 V/m. As for the reflected field, its amplitude is between -40 and -50 dBV/m, i.e. between 0.01 and 0.003 V/m, when it reaches the opposite antenna at 150 m after propagating for 500 ns. Therefore, in this configuration the reflected field is only about two to three orders of magnitude less intense than the incident field.

\begin{figure}
  \includegraphics[width=\linewidth]{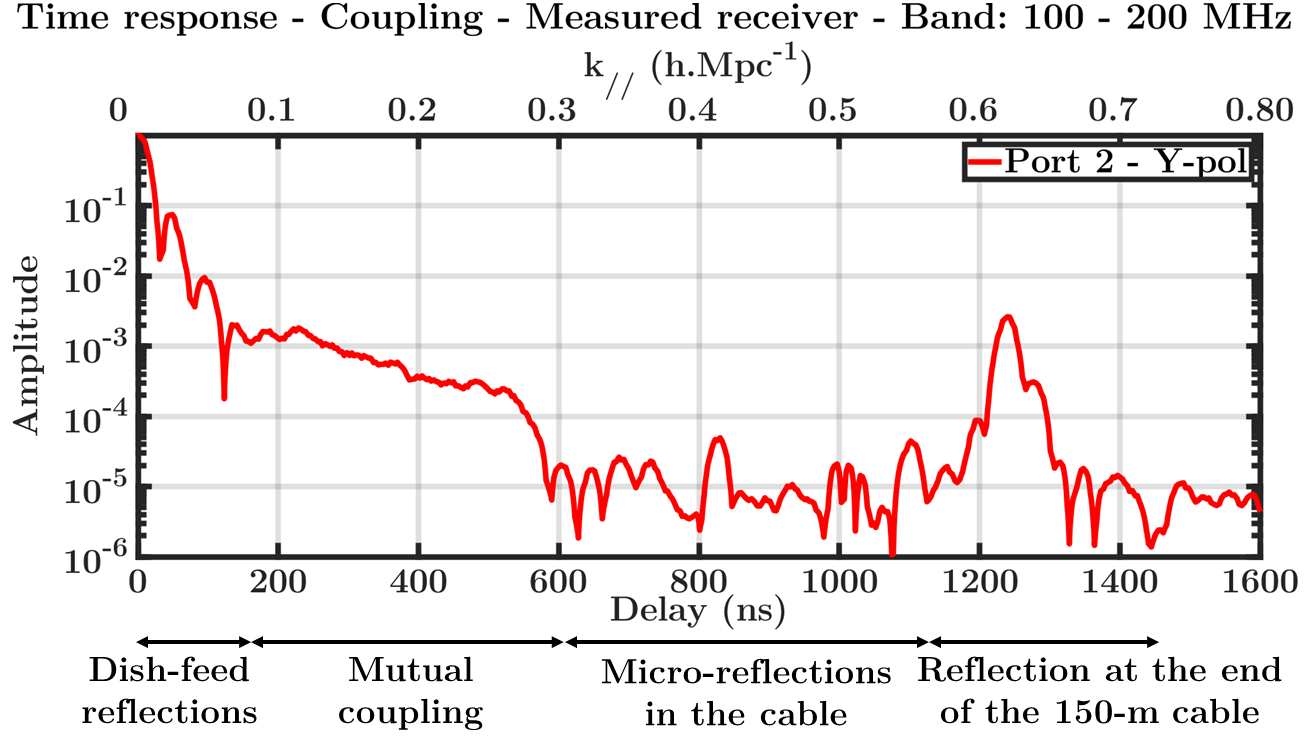}
  \caption{System time response at zenith, terminated by the measured receiver, and including mutual coupling for the port 2 (Y-polarization).}
  \label{fig15:systRespCouplRec}
\end{figure}

\begin{figure}
  \includegraphics[width=\linewidth]{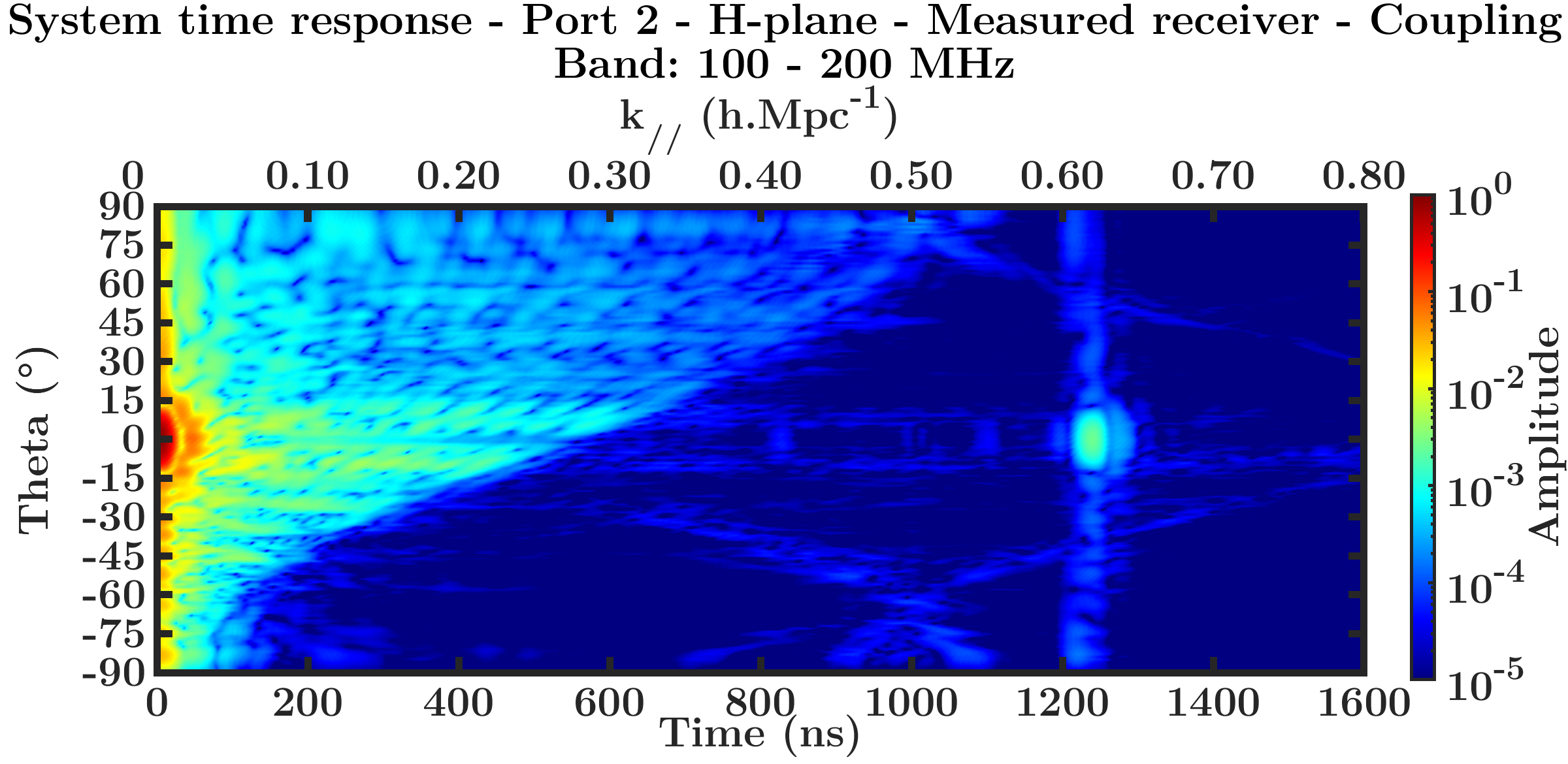}
  \caption{Time response of the system terminated by the measured receiver and including mutual coupling for the port 2 (Y-polarization), and for a plane wave coming with an incidence angle $\theta$ in the H-plane.}
  \label{fig16:systResTimeAngCoupl}
\end{figure}

\begin{figure}
  \includegraphics[width=\linewidth]{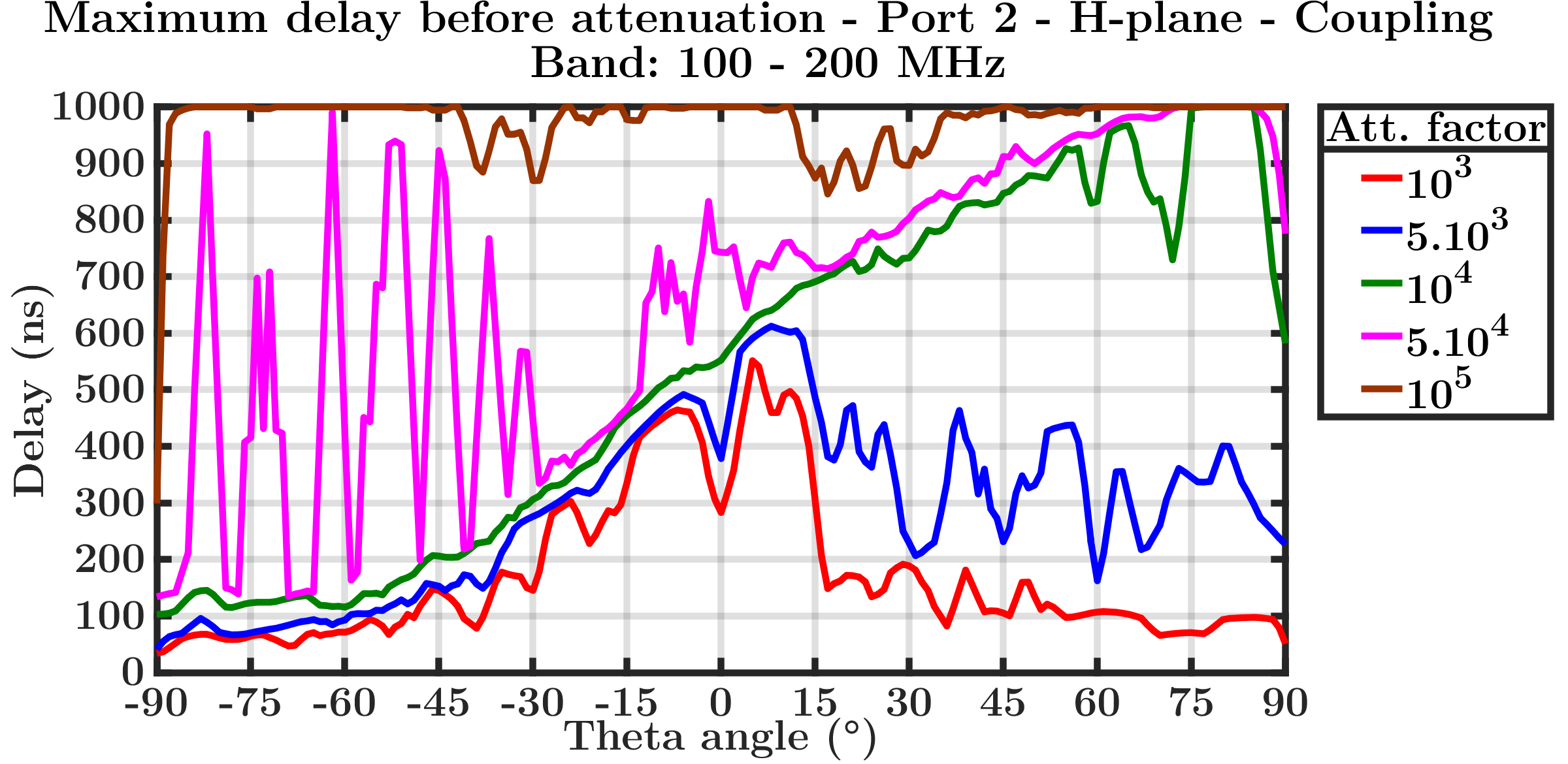}
  \caption{Maximum delays before the system response is attenuated by a certain factor, with the measured receiver, including mutual coupling, for the port 2 (Y-polarization), and for the H-plane.}
  \label{fig17:maxDelCoupl}
\end{figure}

\medskip All the sources of chromaticity are now combined together in a final co-simulation, with the measured receiver and a chromatic cable. The example of the antenna at the edge of the simulated array, port 2, is studied. Fig. \ref{fig15:systRespCouplRec} shows that the system time response can be divided into four distinct parts, each one being dominated by a particular type of chromatic effect: first the feed-vertex reflections, then the mutual coupling, the micro-reflections within the chromatic cable, and the intense reflection at its end. Overall, in a large array the mutual coupling becomes the dominant effect at high delays. Fig. \ref{fig16:systResTimeAngCoupl} and \ref{fig17:maxDelCoupl} illustrate the effects of the incidence angle on the system response in the H-plane. They also show how the spatial and frequency structures of the beam caused by the mutual coupling (cf. Fig. \ref{fig11:chromaBeamCoupl3D}) translate into the time domain. Compared with Fig. \ref{fig10:maxDel}, these plots reveal a significant level of chromaticity for a source near the horizon. When the signal comes with a positive incidence angle in the H-plane, it is first received by the port 2 of the reference antenna, and then propagates through the rest of the array. After a certain geometric delay which depends on the antenna position and on the incidence angle, each antenna receives the main signal which is subsequently reflected back by their dish. Thus, in this simulation the time response spreads up to a characteristic delay ${\tau_{\rm maxCoup}}$ which is equal to the maximum geometric delay associated with the farthest aligned antenna, plus the propagation delay taken by the reflected signal to come back to the reference antenna.
\begin{align}
    {\tau_{\rm maxCoup}} = \frac{{{D_{\rm max}}}}{c}(\sin \left( \theta  \right) + 1),
	\label{eq17:couplDel}
\end{align}
\newline
with ${D_{\rm max}}$ the distance between the reference antenna and the farthest aligned antenna. Thus, for a signal coming from the horizon, the system response of an antenna at the edges can spread up to a delay equal to twice the propagation delay associated with the size of the array. In the same way, if a wave comes from the most populated side of the array, a reference antenna at the edges will receive the main signal approximately at the same time as the contributions from the other antennas. That is why the system response of the port 2 in the H-plane is significantly less chromatic for negative incidence angles. Consequently, the system response of an antenna strongly depends on its position in the HERA array, the incidence angle of the signal, and the considered polarization.

%

%%%%%%%%%%%%%%%%%%%% CONCLUSION %%%%%%%%%%%%%%%%%%

\section{Conclusions on the EoR detection with the foreground avoidance method}
\label{sec:Conclu}

\medskip In this paper, we have presented new models to quantify the chromatic effects in HERA affecting the detection of the EoR delay power spectrum. The success of this experiment relies on our detailed understanding of the instrument. The voltage time response of the system needs to be attenuated by a factor $10^5$ as quick as possible, so it is crucial to have accurate simulations. Our approach aims to be exhaustive and includes the effects of the receiver with the coaxial cable, and the mutual coupling. We have emphasized the importance of combining together all these elements to properly characterize the receiver noise temperature, and the direction-dependent system response in the frequency and time domains. These models are essential to mitigate the chromatic effects, and are used to improve the calibration and the data analysis pipeline \citep{Kern2020b}.

\medskip Previous works based on simulations and measurements \citep{Ewall-Wice2016, Thyagarajan2016, Patra2018} but performed without the receiver parameters and mutual coupling concluded that the system time response would be quickly attenuated by a factor $10^{5}$, and that $k$-modes with a ${k_\parallel}$-component above 0.2 $h\;\rm{Mpc}^{-1}$ should be detectable. This is in agreement with our simulations. However when these elements are included, this new analysis suggests that the detection of the EoR signal using a strict foreground avoidance method will be more challenging than expected with HERA Phase I. The response is attenuated by a factor $10^{5}$ only after 1400 ns, which means that the ${k_\parallel}$-modes below 0.7 $h\;\rm{Mpc}^{-1}$ are affected by the foreground leakage. The time response slowly decays due to the fact that any antenna can interact with the antennas aligned up to the edges of the core. Thus, the attenuation of the response depends on the size of the array and the position of the antenna. This could be problematic to calibrate the system, because this affects the number of truly redundant baselines. The response is also particularly vulnerable to strong radio emission coming from the horizon. Lastly, the signal is significantly reflected at the end of the 150-m cable, and imperfections in this transmission line can also cause micro-reflections all along.

\medskip The analysis of the data obtained with 46 antennas closely packed are consistent with these simulations. These results are detailed in \citet{Kern2020a}. In particular, the system response is studied thanks to the 
interferometric visibility computed in the delay domain (cf. Equation \eqref{eq2:delSpectrum}) and obtained from the auto-correlation of the signal from an antenna. The normalized measured response does show the presence of similar reflections at a comparable level up to 500 ns, as well as the reflections occurring at the end of the cable. However, after 500 ns the amplitude of the response reaches a noise floor at about $10^{-4}$. This can be explained by the fact that the data were collected in a larger two-dimensional array with a more complex configuration compared with our simulation. The cables may also be more chromatic than the one we measured in our laboratory, because of their deterioration in the Karoo desert.

\medskip However, methods based on absolute calibration strategies \citep{Kern2020b} and inverse covariance weighting techniques, such as the "optimal quadratic estimator" (OQE) formalism described in \citet{Liu2010, Dillon2014, Liu2014, Ali2015}, have been developed to try to mitigate the foreground contribution. Thanks to the lessons learnt with this analysis, the system has been re-designed for HERA Phase II. The dipoles are being replaced by Vivaldi feeds \citep{Fagnoni2020}, and new receivers are being deployed. These new feeds have a wider bandwidth from 50 to 250 MHz, and are also slightly less chromatic since they do not require a cage to properly radiate towards the dish. The FEM and PAM are connected together using 500-m optical fibres, which should solve the problems related to the coaxial cables, and unlock lower ${k_\parallel}$-modes. In this context, mutual coupling becomes the limiting factor. The size of the array is an advantage to reach the sensitivity required to detect the EoR signal, but it also increases the level of chromaticity. Therefore, new solutions to limit the scattering and the propagation of the signal from one dish to another one need to be studied, for instance by treating the antenna rims with a resistive surface \citep{Bucci1981, Jenn1991}.

%%%%%%%%%%%%%%%%%%%% ACKNOWLEDGMENT %%%%%%%%%%%%%%%%%%

\section*{Acknowledgements}

This material is based upon work supported by the National Science Foundation under Grant Nos. 1636646 and 1836019 and institutional support from the HERA collaboration partners. The authors also acknowledge the UK Science and Technology Facilities Council (STFC), as well as the Institute of Astronomy and the Physics Department of the University of Cambridge (UK) for their financial support via the Isaac Newton Studentship.

%%%%%%%%%%%%%%%%%%%% DATA AVAILABILITY %%%%%%%%%%%%%%%%%%

\section*{DATA AVAILABILITY}

The data underlying this article will be shared on reasonable request to the corresponding author.



\bibliographystyle{mnras}
\bibliography{Co-simulations_of_the_HERA_Phase_I_receiver_system-MNRAS-resubmit-v3} %Biblio.bib

%\begin{thebibliography}{99}
%\end{thebibliography}

%%%%%%%%%%%%%%%%%%%%%%%%%%%%%%%%%%%%%%%%%%%%%%%%%%
% Don't change these lines
\bsp	% typesetting comment
\label{lastpage}
\end{document}